\documentclass[journal,comsoc]{IEEEtran}
\usepackage[T1]{fontenc}

\ifCLASSINFOpdf
  \usepackage{threeparttable}
  \usepackage[justification=centering]{caption}
  \usepackage[linesnumbered,ruled,vlined]{algorithm2e}
  \usepackage{amsmath}
  \usepackage{epsfig,amssymb,subfigure,bm,dsfont}
  \usepackage{algpseudocode}
  \usepackage{amsfonts}
  \usepackage{array}
  \usepackage{balance}
  \usepackage{bbding}
  \usepackage{booktabs}
  \usepackage{calc}
  \usepackage{cite}
  \usepackage{citesort}
  \usepackage{color}
  \usepackage{curves}
  \usepackage{epstopdf}
  \usepackage{epic}
  \usepackage{float}
  \usepackage{footnote}
  \usepackage{graphicx}
  \usepackage{makecell}
  \usepackage{mdwlist}
  \usepackage{multicol}
  \usepackage{multirow}
  \usepackage{ntheorem}
  \usepackage{pifont}
  \usepackage[english]{babel}
  \usepackage{subfig}
  \usepackage{graphicx} 
  \usepackage{epstopdf}
\else
  \usepackage{cite}
\fi

\usepackage[cmintegrals]{newtxmath}

\begin{document}
%

\title{Collaborative Intelligent Reflecting Surface Networks with Multi-Agent Reinforcement Learning}

\author{Jie Zhang,
        Jun Li,
        Yijin Zhang,
        Qingqing Wu,
        Xiongwei Wu,
        Feng Shu,
        Shi Jin,
        Wen Chen
        \IEEEcompsocitemizethanks{
        \IEEEcompsocthanksitem This work was supported in part by National Natural Science Foundation of China under Grants 61872184, 62071234, 62071236 and 62071296, in part by National Key Project under Grants 2018YFB1801102 and 2020YFB1807700, in part by Shanghai Fundamental Project under Grant 20JC1416502, in part by the Hainan Major Projects ZDKJ2021022, in part by the Scientific Research Fund Project of Hainan University under Grant KYQD(ZR)-21008, and in part by the Fundamental Research Funds for the Central Universities of China No. 30921013104 and No. 30920021127. ($\emph{Corresponding author: Jun Li}$).
        \IEEEcompsocthanksitem J. Zhang, J. Li and Y. Zhang are with the School of Electronic and Optical Engineering, Nanjing University of Science and Technology, 210094, China (e-mail: \{zhangjie666; jun.li\}@njust.edu.cn; yijin.zhang@gmail.com).
        \IEEEcompsocthanksitem Q. Wu is with the State Key Laboratory of IoT for Smart City, University of Macau, Macao 999078, China (email: qingqingwu@um.edu.mo).
        \IEEEcompsocthanksitem X. Wu is with the Department of Electronic Engineering, The Chinese University of Hong Kong, Hong Kong SAR of China (e-mail: xwwu@ee.cuhk.edu.hk).
        \IEEEcompsocthanksitem F. Shu is with the School of Information and Communication Engineering, Hainan University, Haikou 570228, China (email: shufeng0101@163.com).
        \IEEEcompsocthanksitem S. Jin is with the National Mobile Communications Research Laboratory, Southeast University, Nanjing 210096, China (e-mail: jinshi@seu.edu.cn).
        \IEEEcompsocthanksitem W. Chen is with the Department of Electronic Engineering, Shanghai Jiao Tong University, Shanghai 200240, China (e-mail: wenchen@sjtu.edu.cn).
        }
        }

\markboth{IEEE Journal of selected topics in signal processing,~Vol.~XX, No.~XX,~2021}%
{Shell \MakeLowercase{\textit{et al.}}: Bare Demo of IEEEtran.cls for IEEE Communications Society Journals}

\maketitle

\begin{abstract}
Intelligent reflecting surface (IRS) is envisioned to be widely applied in future wireless networks. In this paper, we investigate a multi-user communication system assisted by cooperative IRS devices with the capability of energy harvesting. Aiming to maximize the long-term average achievable system rate, an optimization problem is formulated by jointly designing the transmit beamforming at the base station (BS) and discrete phase shift beamforming at the IRSs, with the constraints on transmit power, user data rate requirement and IRS energy buffer size. Considering time-varying channels and stochastic arrivals of energy harvested by the IRSs, we first formulate the problem as a Markov decision process (MDP) and then develop a novel multi-agent Q-mix (MAQ) framework with two layers to decouple the optimization parameters. The higher layer is for optimizing phase shift resolutions, and the lower one is for phase shift beamforming and power allocation. Since the phase shift optimization is an integer programming problem with a large-scale action space, we improve MAQ by incorporating the Wolpertinger method, namely, MAQ-WP algorithm to achieve a sub-optimality with reduced dimensions of action space. In addition, as MAQ-WP is still of high complexity to achieve good performance, we propose a policy gradient-based MAQ algorithm, namely, MAQ-PG, by mapping the discrete phase shift actions into a continuous space at the cost of a slight performance loss. Simulation results demonstrate that the proposed MAQ-WP and MAQ-PG algorithms can converge faster and achieve data rate improvements of $10.7\%$ and $8.8\%$ over the conventional multi-agent DDPG, respectively. 
\end{abstract}


%
\IEEEpeerreviewmaketitle

\section{Introduction}
%
%
%
%
\IEEEPARstart{R}{encently}, advanced technologies have been developed in the fifth generation (5G), such as massive multiple input multiple output (MIMO) and network densification~\cite{5G1,5G2}, to achieve high throughput, ultra low latency and high reliability. However, progressively dense deployment of MIMO base stations (BSs) with large-scale multi-antenna arrays suffers high cost and substantial power consumption.

To tackle this challenge, a promising paradigm called intelligent reflecting surface (IRS) has been proposed and has aroused considerable research enthusiasm due to its superiority of energy efficiency and low cost~\cite{8910627}. An IRS can be seen as a flat composed of many passive reflecting units, which is able to tune the phases and amplitudes of incident signals and then reflect them into the desired directions. By this way, it can significantly enhance the signal-to-interference-plus-noise ratio (SINR)~\cite{8936989}. Notably, an IRS can reflect the input signals passively by controlling the electronic devices without the need of utilizing radio-frequency chains. In addition, an IRS is generally portable and scalable, and thus can be easily deployed on the indoor furnitures and outdoor walls~\cite{IRSTutorial}. Owing to these nice features, the IRS technology is expected to be widely applied in various wireless communication scenarios to improve the network performance.

\subsection{Related Work}
Many recent studies have been devoted to investigating properties and challenges of the IRS-assisted conventional systems. Such a system with a single-antenna BS and multiple single-antenna users was studied in \cite{8811733}. For IRS-assisted multi-antenna systems, \cite{8743496,8723525,9115725} employed an alternating optimization (AO) approach and a semi-definite relaxation (SDR) algorithm to maximize the signal-to-noise ratio (SNR) with a guarantee of the user secrecy rate. Zhou \emph{et al}.~\cite{PanIRS} investigated the robust beamforming at the BS for IRS-assisted multiple-input single-output (MISO) systems. For an IRS-assisted multi-user MIMO (MU-MIMO) system, the study in~\cite{8941126} used a two-step stochastic program to formulate the average received SNR maximization issue and utilized a minorization-maximization (MM) based algorithm to solve the passive beamforming and information transfer problem.

In addition, the IRS technology has been applied to some novel communication systems. For an IRS-assisted millimeter-wave system, Xiu \emph{et al}.~\cite{9419998} designed the IRS beamforming to offer more feasible propagation paths, and proposed an alternating manifold optimization method to maximize the weighted sum rate. For an IRS-assisted unmanned aerial vehicle (UAV) system, the studies in~\cite{9120632,9400768,9078125} investigated how to increase the received signal strength by passive beamforming at each IRS. For a simultaneous wireless information and power transfer (SWIPT) system, the studies in~\cite{8941080,IRSSWIPT,IRSSWIPT2} employed IRSs to serve energy harvesting receivers and information decoding receivers, and optimized the IRSs phase shifts to maximize the weighted sum-power. For an IRS-enhanced orthogonal frequency division multiplexing (OFDM) system, Yang \emph{et al}.~\cite{9039554} maximized the achievable rate through reasonably allocating the transmit power and designing the passive beamforming.

Regarding the design of IRS passive beamforming, most previous studies concentrated on the continuous phase shift optimization at each IRS, which leads to excessively high resolution, and thus its computational burden is unacceptable in practice. When each IRS has a finite number of phase shifts, Wu \emph{et al}.~\cite{8930608} demonstrated that the discrete phase shifts and continuous phase shifts can achieve the same power gain. Further, the study in~\cite{9133142} designed the IRS passive beamforming under discrete phase shifts and imperfect channel state information. 
However, the above mentioned works are not applicable to dynamic environments (e.g., varying channels, stochastic arrivals of mobile users) by utilizing traditional optimization approaches, e.g., AO, SDR, MM algorithms, to address the beamforming optimization problems. 

Artificial intelligence (AI) has recently developed as a remarkably impressive technology to tackle dynamic optimization problems in large-scale systems~\cite{9069945,You2019AI,9446746,QianUAVRL,ChuanFL}. Considering the superiority of AI, deep learning (DL) has been used to maximize the user received signal strength by formulating the IRSs online wireless configuration~\cite{HuangDL}. The studies in~\cite{8968350,IRSRL} applied reinforcement learning (RL) to achieve the maximum SNR by optimizing the IRS passive phase shift. The work in~\cite{9206080,IRSRLenergy} developed deep RL (DRL) methods to improve the system secrecy rate and the energy efficiency by jointly optimizing the BS beamforming and the IRSs' reflecting beamforming. In~\cite{9410457,IRSRLMISO}, the joint design of the BS digital beamforming and the IRSs' analog beamforming was formulated as an NP-hard optimization problem to improve the coverage range by leveraging DRL. 

To the best of our knowledge, multi-agent RL (MARL) algorithm has not been developed in the existing works to cope with the joint beamforming optimization problems in multiple IRSs-assisted multi-user systems, under the condition of time-varying discrete phase shifts design and energy harvesting mechanism.

\subsection{Contributions}
In this paper, we aim to maximize the long-term average achievable system data rate by optimizing BS transmit beamforming and IRSs' discrete phase shift beamforming with transmit power limits, user data rate requirements and IRS energy storage buffer constraints, assuming that each IRS has adjustable phase shift resolutions and is equipped with energy harvesting devices. The main contributions of this paper are summarized as follows:
\begin{itemize}
\item [1)] 
Considering a multi-user MISO system assisted by distributed IRSs and a central BS, we formulate the joint transmit beamforming and phase shift beamforming optimization with the objective of maximizing the long-term average achievable system rate and propose a cooperative multi-agent Markov decision process (MDP) to model the distributed IRS-assisted system due to the time-varying channels and the stochastic harvested energy. The BS and all the IRSs are considered as agents that can interact with the 
system environment and learn through the historical interactive experience.
\item [2)]
We develop a novel multi-agent Q-mix (MAQ) framework with two layers to decouple the optimization parameters. The high-level layer is for optimizing phase shift resolutions and the low-level layer is for phase shift beamforming and power allocation. To efficiently handle the exponentially large number of phase shift actions, we propose a MAQ with Wolpertinger method (MAQ-WP) algorithm to obtain sub-optimal beamforming policies. We further propose a MAQ with policy gradient (MAQ-PG) algorithm to overcome the weakness of high complexity in the MAQ-WP algorithm at the cost of a slight performance loss.
\item [3)] 
We generalize the MAQ-WP and MAQ-PG algorithms by segregating the high-dimensional and discrete phase shift actions into two hierarchical actions to significantly accelerate the learning process and reduce the computational complexity. We propose Wolpertinger and policy gradient methods to map proto-actions with high dimensionality into actual actions with low dimensionality, thereby improving the learning rate effectively.
\item [4)] 
Extensive simulation results demonstrate the effectiveness of the proposed algorithms in improving both the convergence value and learning speed under the constraints compared with benchmarks. 
In addition, the MAQ-WP and MAQ-PG algorithms can increase the long-term average achievable system rate by $10.7\%$ and $8.8\%$ compared with the multi-agent deep deterministic policy gradient (MADDPG)~\cite{MADDPG} based approach, respectively. 
\end{itemize}

The rest of this paper is organized as follows. The system model and the problem formulation are provided in Section~\ref{sec:System Model}. The MAQ framework based on an MDP formulation and the proposed MAQ-WP and MAQ-PG algorithms are presented in Section~\ref{sec:DRL-BASED SOLUTION}. Section~\ref{sec:SIMULATION RESULTS} provides numerical results to evaluate the proposed algorithms. Section~\ref{sec:CONCLUSIONS} concludes this paper.

\emph{Math notation}: Vectors and matrices are denoted by boldface lowercase letters $\boldsymbol{a}$ and boldface capital letters $\boldsymbol{A}$, respectively. $(\cdot)^{T}$ and $(\cdot)^{H}$ are transpose and conjugate transpose operations, respectively. Let $\left|\cdot\right|$ and $\left\|\cdot\right\|$ denote the absolute value and the Euclidean norm operations, respectively. The operator $\rm diag(\cdot)$ represents the diagonal matrix with argument of a vector. We use $x\sim\mathcal{CN}(0,1)$ to indicate that the random variable $x$ obeys the complex Gaussian distribution with zero-mean and unit variance.

\section{System Model}\label{sec:System Model}
In this section we introduce the considered signal model and energy harvesting model. Based on these two models, under the objective of maximizing the long-term average system data, we formulate a constrained optimization problem. For ease of reference, Table \ref{tab:sum_nota} lists all the main notation of the system model.
\begin{table}[ht]
\caption{Summary of main notation}
\centering
\begin{tabular}{l l}
\hline
\makecell[c]{\textbf{Notation}} & \makecell[c]{\textbf{Description}} \\
\hline
\makecell[c]{$L$,$\mathcal{L}$} & \makecell[c]{Number of IRSs, set of IRSs}\\
\makecell[c]{$K$,$\mathcal{K}$} & \makecell[c]{Number of users, set of users}\\
\makecell[c]{$\textbf{h}_{l,k}^{\mathrm{RU}}$} & \makecell[c]{Channel state information between $l$-th IRS and $k$-th user}\\
\specialrule{0em}{1pt}{1pt}
\makecell[c]{$\textbf{h}_{k}^{\mathrm{BU}}$} & \makecell[c]{Channel state information between BS and the $k$-th user}\\
\specialrule{0em}{1pt}{1pt}
\makecell[c]{$\textbf{H}_{l}^{\mathrm{BR}}$} & \makecell[c]{Channel state information between BS and the $l$-th IRS}\\
\makecell[c]{$\boldsymbol{\Phi}_{l}$} & \makecell[c]{The reflection matrix of the $l$-th IRS}\\
\makecell[c]{$b_{l}$} & \makecell[c]{Bit resolution of the $l$-th IRS}\\
\makecell[c]{$\rho_{l,n}(t)$} & \makecell[c]{Working status of the $l$-th IRS $n$-th element at the $t$-th slot}\\
\makecell[c]{$P^{\mathrm{IRS}}(t)$} & \makecell[c]{total power consumption of all the IRSs at the $t$-th slot}\\
\makecell[c]{$E_{l}(t)$} & \makecell[c]{Remaining energy of the $l$-th IRS at the $t$-th slot}\\
\hline
\end{tabular}
\label{tab:sum_nota}
\end{table}

\subsection{Signal Model}
\begin{figure}[ht]
    \centering
    \includegraphics[width=7cm,height=7cm]{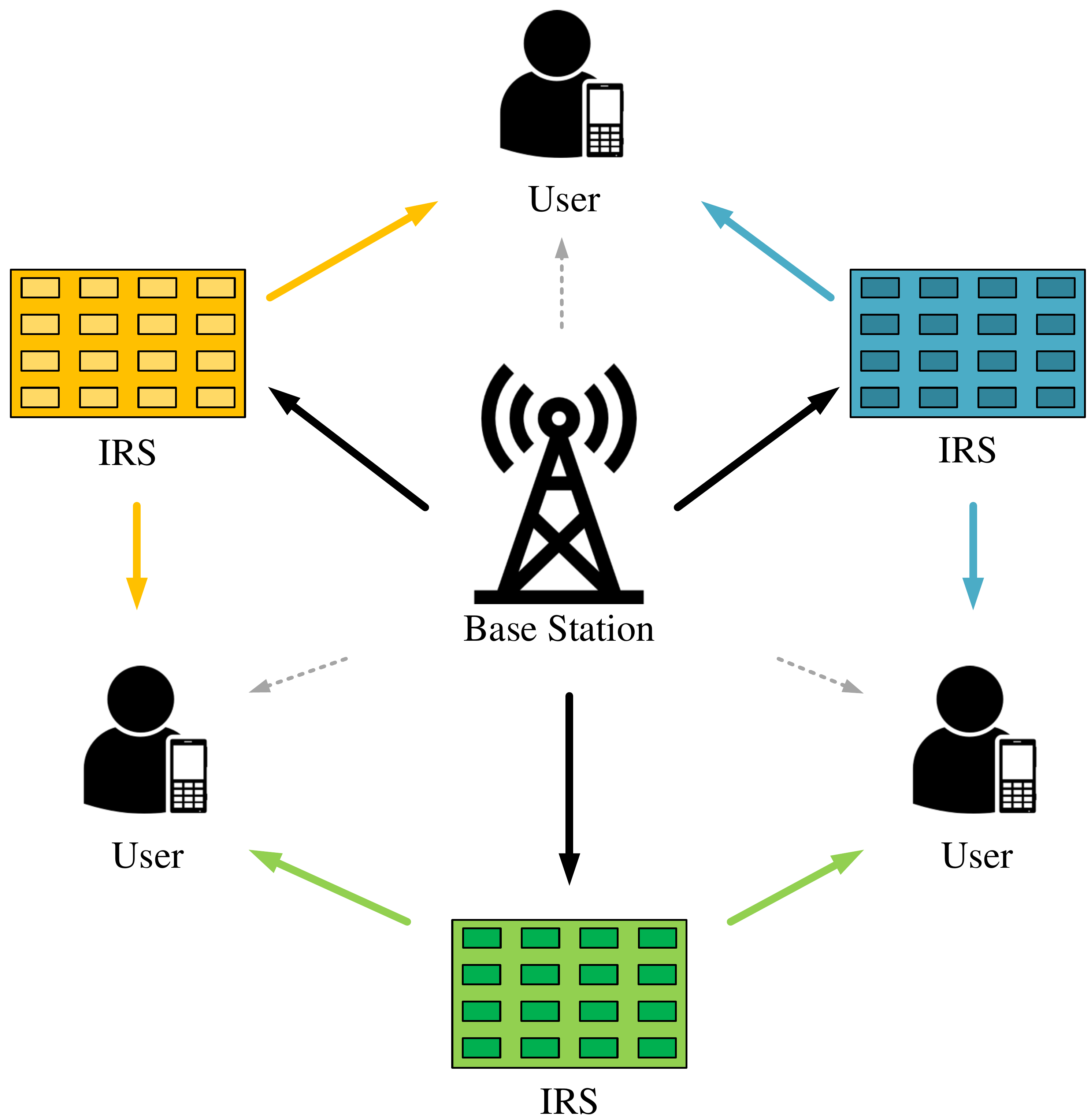}\\
    \caption{Downlink multi-user MISO system with multiple IRSs}
    \label{system model picture}
\end{figure}

As illustrated in Fig.~\ref{system model picture}, we consider a multi-user MISO communication system assisted by multiple IRSs, where a downlink BS equipped with $M$ antennas communicates with $K$ single-antenna users under $L$ IRSs' cooperation. Let $\mathcal{K}\triangleq \{1,...,K\}$ and $\mathcal{L}\triangleq \{1,...,L\}$ denote indices of users and indices of IRSs, respectively. It is assumed that the IRS $l,\forall l\in\mathcal{L}$ is equipped with $N_{l}$ reflecting elements or unit cells. We denote the $l$-th IRS reflecting elements set as $\mathcal{N}_{l}\triangleq \{1,...,N_{l}\}$.\\
\; The received signal at the $k$-th user can be given by
\begin{equation}\label{received signal 1}
y_{k}=\left(\sum_{l=1}^{L}\left(\textbf{h}_{l,k}^{\mathrm{RU}}\right)^{H}\boldsymbol{\Phi}_{l}\textbf{H}_{l}^{\mathrm{BR}}+\left(\textbf{h}_{k}^{\mathrm{BU}}\right)^{H}\right)\sum_{j=1}^{K}\textbf{v}_{j}s_{j}+n_{k},
\end{equation}
where $\textbf{h}_{l,k}^{\mathrm{RU}}\in\mathbb{C}^{N_{l}\times1}$ denotes the channel coefficients vector between the $l$-th IRS and the $k$-th user, $\textbf{H}_{l}^{\mathrm{BR}}\in\mathbb{C}^{N_{l}\times M}$ denotes the channel coefficients matrix between the BS and the $l$-th IRS, $\textbf{h}_{k}^{\mathrm{BU}}\in\mathbb{C}^{M\times1}$ denotes the channel coefficients vector between the BS and the $k$-th user. $\boldsymbol{\Phi}_{l}\triangleq\mathrm {diag}(\boldsymbol{\phi}_{l})\in\mathbb{C}^{N_{l}\times N_{l}}$ denotes reflection coefficients matrix of the $l$-th IRS, with the $l$-th IRS reflection coefficients vector $\boldsymbol{\phi}_{l}$ being defined as
\begin{equation}\label{reflection vector}
\boldsymbol{\phi}_{l}\triangleq\left[\beta_{l,1}e^{j \theta_{l,1}},\beta_{l,2}e^{j \theta_{l,2}},...,\beta_{l,N_{l}}e^{j \theta_{l,N_{l}}}\right]^{T},
\end{equation}
where $\beta_{l,i} \in [0,1],i\in\mathcal{N}_{l}$ denotes the $i$-th element amplitude reflection coefficient of the $l$-th IRS, and $\theta_{l,i} \in [0,2\pi),i\in\mathcal{N}_{l}$ denotes the $i$-th element phase shift reflection coefficient of the $l$-th IRS. 
In this paper, we assume that the amplitude reflection coefficient of each IRS element is set to be one for maximizing the signal reflection, i.e., $\beta_{l,i}=1,\forall i\in\mathcal{N}_{l}, l\in\mathcal{L}$. The transmit beamforming vector $\textbf{v}_{j}\in\mathbb{C}^{M\times 1},\forall j\in\mathcal{K}$ and information symbol $s_{j}\sim\mathcal{C}\mathcal{N}(0,1),\forall j\in\mathcal{K}$ are designed for the $j$-th user. Let $n_{k}\sim\mathcal{CN}(0,\sigma_{n}^{2})$ denote the additive white Gaussian noise (AWGN) with zero mean and $\sigma_{n}^{2}$ variance. We separate the $k$-th user's received signal expressed in (\ref{received signal 1}) into three parts: desired signal, inter-user interference signal and noise signal, i.e.,
\begin{equation}\label{received signal 2}
\begin{split}
\begin{aligned}
y_{k}=&\underbrace{\left(\sum_{l=1}^{L}\left(\textbf{h}_{l,k}^{\mathrm{RU}}\right)^{H}\boldsymbol{\Phi}_{l}\textbf{H}_{l}^{\mathrm{BR}}+\left(\textbf{h}_{k}^{\mathrm{BU}}\right)^{H}\right)\textbf{v}_{k}s_{k}}_{\text{desired signal}}+\\
&\underbrace{\sum_{j,j\neq k}^{K}\left(\sum_{l=1}^{L}\left(\textbf{h}_{l,k}^{\mathrm{RU}}\right)^{H}\boldsymbol{\Phi}_{l}\textbf{H}_{l}^{\mathrm{BR}}+\left(\textbf{h}_{k}^{\mathrm{BU}}\right)^{H}\right)\textbf{v}_{j}s_{j}}_{\text{inter-user interference signal}}+\underbrace{n_{k}}_{\text{noise signal}}.
\end{aligned}
\end{split}
\end{equation}

Then the SINR at the $k$-th user can be obtained by
\begin{equation}\label{SINR}
\gamma_{k}=\frac{\left|\left(\sum_{l=1}^{L}\left(\textbf{h}_{l,k}^{\mathrm{RU}}\right)^{H}\boldsymbol{\Phi}_{l}\textbf{H}_{l}^{\mathrm{BR}}+\left(\textbf{h}_{k}^{\mathrm{BU}}\right)^{H}\right)\textbf{v}_{k}\right|^{2}}{\left|\sum_{j,j\neq k}^{K}\left(\sum_{l=1}^{L}\left(\textbf{h}_{l,k}^{\mathrm{RU}}\right)^{H}\boldsymbol{\Phi}_{l}\textbf{H}_{l}^{\mathrm{BR}}+\left(\textbf{h}_{k}^{\mathrm{BU}}\right)^{H}\right)\textbf{v}_{j}\right|^{2}+\sigma_{n}^{2}},
\end{equation}
and thus, the achievable system data rate can be obtained as
\begin{equation}\label{system achievable rate}
R=\sum_{k=1}^{K}\log_{2}\left(1+\gamma_{k}\right).
\end{equation}

\subsection{Energy Harvesting Model}
An IRS is a passive component that reflects incident signal without amplification in theory. However, an IRS indeed consumes energy in practice due to the operations of IRS elements, IRS controller and circuit board. Hence, the IRS power consumption depends on the phase shift resolution and the operation status per IRS element.

Ideally, the phase shift of each IRS element can be tuned continuously. In practice, however, the phase shift is finite and discrete due to the complex hardware limitation. Let $\mathcal{B}\triangleq\{1,...,B\}$ denote the set of all possible bit resolution values. Then the set of all possible discrete phase shifts taking the $b$-bit resolution can be indicated as
\begin{equation}\label{phase shift set}
\mathcal{F}_{b}\triangleq\{0,\Delta \theta,...,(2^{b}-1)\Delta \theta\},\forall b\in\mathcal{B},
\end{equation}
where $\Delta \theta = 2\pi/2^{b}$ and thus each element is within the range of $[0,2\pi)$. We assume that each IRS can alter its phase shift bit resolution according to the actual requirement. For example, as shown in~\cite{8741198}, power consumption per element is 1.5, 4.5 and 6mW for 3-,4- and 5-bit resolution phase shifting, respectively. The higher the resolution of IRS phase shifts, the better its beamforming performance, but the power consumption will increase greatly. Therefore, there is a tradeoff between resolution and power consumption. Such power consumption is much lower than that of an amplify-and-forward (AF) relay. However, in a large-scale multiple IRSs system, the IRS power consumption is still considerable.

In our work, a time-slotted system with a minimum slot length $\Delta t$ is considered. Each IRS is assumed to carry an energy storage buffer and an energy harvesting device, which can gather solar energy with the usage of solar panels. It is assumed that the harvested energy obeys a certain statistical distribution, e.g., Poisson distribution. By fabricating the tunable reflecting element based on PIN diode, we can set each IRS element to an ON or OFF working status. If an IRS element works at the ON status, the incident signal will be reflected by the IRS element without amplitude attenuation and the signal phase will be changed. Otherwise, the incident signal will not be reflected and the signal power will not be absorbed into the IRS energy storage buffer. Thus, by adjusting the working status of the IRSs elements, the desired signal can be boosted and the interference signals can be weakened effectively. 

The status of the $l$-th IRS $n$-th element at the $t$-th slot is denoted as $\rho_{l,n}(t)$. If $\rho_{l,n}(t)=1$, the element operates at the ON status at the $t$-th slot. We assume that all elements in the same IRS have the same phase shift resolution. The ON-status power consumption of the $l$-th IRS element under $b_{l}\in\mathcal{B}$ bit resolution at the $t$-th slot is denoted as $\mu_{l}(t)$. As a result, the power consumption of the $l$-th IRS at the $t$-th slot can be expressed as
\begin{equation}\label{each IRS power consumption}
P_{l}^{\mathrm{IRS}}(t)=\sum_{n=1}^{N_{l}}\rho_{l,n}(t)\mu_{l}(t),\forall l\in\mathcal{L}.
\end{equation}

Then total power consumption of all the IRSs at the $t$-th slot can be obtained as
\begin{equation}\label{IRS power consumption}
P^{\mathrm{IRS}}(t)=\sum_{l=1}^{L}P_{l}^{\mathrm{IRS}}(t).
\end{equation}

The remaining energy stored in the $l$-th IRS energy storage buffer can be updated by
\begin{equation}\label{remaining energy}
E_{l}(t+1) \triangleq \min\left\{\max\left\{E_{l}(t)-c_{l}(t),E_{l}^{\mathrm{min}}\right\}+a_{l}(t),E_{l}^{\mathrm{max}}\right\},
\end{equation}
where $E_{l}(t)$ is the remaining stored energy of the $l$-th IRS at the $t$-th slot, $c_{l}(t)=P_{l}^{\mathrm{IRS}}(t)\Delta t$ is the energy consumption of the $l$-th IRS at the $t$-th slot, $a_{l}(t)$ is the harvested energy at the $t$-th slot, $E_{l}^{\mathrm{min}}$ and $E_{l}^{\mathrm{max}}$ are the minimum threshold and the maximum capacity of the energy storage buffer, respectively. 


\subsection{Problem Formulation}
Our main objective is to jointly design the BS transmit beamforming matrix $\textbf{V}\triangleq\left[\textbf{v}_{1},...,\textbf{v}_{K}\right]$, the IRS element bit resolution vector $\textbf{b}\triangleq\left[b_{1},...,b_{L}\right]$, the IRS reflecting matrix $\boldsymbol{\Phi}_{l},\forall l\in\mathcal{L}$ and the ON/OFF status vector $\boldsymbol{\rho}_{l}\triangleq\left[\rho_{l,1},...,\rho_{l,N_{l}}\right],\forall l\in\mathcal{L}$ to maximize the long-term average achievable system data rate in the multiple IRSs-assisted communication system.

Let $\textbf{X}(t)\triangleq\left[\textbf{V}(t),\textbf{b}(t),\{\boldsymbol{\Phi}_{l}(t)\}_{1}^{L},\{\boldsymbol{\rho}_{l}(t)\}_{1}^{L}\right]$ denote the set consisting of all the variables at the $t$-th slot. Let $\{x_{l}(t)\}_{1}^{L}\triangleq \{x_{1}(t),...,x_{L}(t)\}$ for convenience. We aim to maximize the long-term average achievable system data rate $\lim\limits_{T\to\infty}\frac{1}{T}\sum_{t=0}^{T}R(t)$ subject to the data rate requirement $R_{k}^{\rm req}(t)$ of the $k$-th user at the $t$-th slot, the transmit power limitation $P^{\mathrm{max}}$ at the BS and the energy storage buffer constraints $E_{l}^{\mathrm{min}},E_{l}^{\mathrm{max}}$ at the $l$-th IRS. Accordingly, the optimization function is formulated as
\begin{align}
\textbf{P}:&\max_{\textbf{X}(t)}\lim\limits_{T\to\infty}\frac{1}{T}\sum_{t=0}^{T}R(t),\label{optimization function}\\
&\text{s.t.}\;C1\!:\!\ \sum_{k=1}^{K}P_{k}(t)\leq P^{\mathrm{max}}, \forall t,\tag{\ref{optimization function}{a}} \label{optimization function constraint 1}\\
&\;\,\quad C2\!:\!\!\ R_{k}(t)\geq R_{k}^{\rm req}(t), \forall k\in \mathcal{K},\forall t,\tag{\ref{optimization function}{b}} \label{optimization function constraint 2}\\
&\;\,\quad C3\!:\!\!\ E_{l}^{\mathrm{min}}\leq E_{l}(t) \leq E_{l}^{\mathrm{max}}, \forall l\in \mathcal{L},\forall t\tag{\ref{optimization function}{c}} \label{optimization function constraint 3},
\end{align}
where $P_{k}(t)=\left\|\textbf{v}_{k}\right\|$ is the transmit power of the $k$-th user signal, $R_{k}(t)=\log_{2}\left(1+\gamma_{k}(t)\right)$\footnote{Without further explanation, in the following sections, notation with subscript $t$ will substitute the functional form notation at the $t$-th slot for convenience, e.g., ($R_{k}(t)\rightarrow R_{k,t}$).} and $R(t)$ are the achievable data rate of the $k$-th user and the achievable system data rate at the $t$-th slot, respectively. The optimization function (\ref{optimization function}) computes the long-term average achievable system data rate. The constraint (\ref{optimization function constraint 1}) indicates that the total transmit power is no more than the transmit power constraint $P^{\mathrm{max}}$. The constraint (\ref{optimization function constraint 2}) means that each user achievable data rate $R_{k}(t)$ should satisfy his data rate requirement $R_{k}^{\rm req}(t)$. The constraint (\ref{optimization function constraint 3}) takes into account the IRS energy storage buffer size.

The formulated objective function in (\ref{optimization function}) with constraints in (\ref{optimization function constraint 1}), (\ref{optimization function constraint 2}) and (\ref{optimization function constraint 3}) is a non-convex problem because the transmit beamforming and phase shifts are coupled in~(\ref{SINR}). As conventional optimization techniques (e.g. convex optimization methods) are challenging to solve it, we aim to a DRL-based method to address this problem.

\section{DRL-based Solution}\label{sec:DRL-BASED SOLUTION}
In this section, we construct an MDP formula for the proposed constrained optimization function and then propose an MAQ structure and two algorithms to solve the optimization problem.

\subsection{MDP Formula}
The optimization problem given in (\ref{optimization function}) is a complex non-convex problem with three types of constraint conditions. Besides, the channel state information, the user data rate requirements and the harvested energy are all time-varying. Thus, traditional optimization approaches that transform a
dynamic system into a static system may achieve poor performance and have no guarantee of constraints. As such, we transform the dynamic system into a cooperative multi-agent MDP model by viewing the multiple IRSs-assisted multi-user MISO
communication system in Section~\ref{sec:System Model} as an interactive system and viewing the BS and all the IRSs as distributed learning agents. The main elements of the constructed multi-agent MDP are defined as follows:

\subsubsection{State}\label{state definition}
The local state of the $l$-th IRS at the $t$-th slot should reflect the user data rate and the energy consumption, which is defined as
\begin{equation}\label{IRS state}
\boldsymbol{s}_{l,t}\triangleq\left[\left\{\exp\left({-\frac{d_{l,k}}{d_{l,0}}}\right)\cdot f_{k}\left(R_{k,t-1}\right)]\right\}_{k\in \mathcal{K}}, E_{l,t}\right],
\end{equation}
where $\exp\left({-\frac{d_{l,k}}{d_{l,0}}}\right)$ is a discount factor that embodies the impact of the $k$-th user on the $l$-th IRS, $d_{l,k}$ is the distance between the $l$-th IRS and the $k$-th user, $d_{l,0}$ is the reference distance of the $l$-th IRS. The indicator function $f_{k}(R_{k,t})$ is defined as
\begin{equation}\label{indicator function 1}
f_{k}(R_{k,t})\triangleq\left\{
\begin{aligned}
&1 \qquad \rm if\;\emph{R}_{\emph{k},\emph{t}} > \emph{R}_{\emph{k},\emph{t}}^{\rm req}, \\
&0 \qquad \rm else.
\end{aligned}
\right.
\end{equation}
The local state $s_{l,t}$ indicates that if the user is adjacent to the IRS and its data rate requirement is satisfied, the user will be significant to the IRS. The local state of the BS at the $t$-th slot is defined as the transmit power, i.e.,
\begin{equation}\label{BS state}
\boldsymbol{s}_{\rm B,\emph{t}}\triangleq\left[\left\{P_{k,t}\right\}_{k \in \mathcal{K}}\right].
\end{equation}

The local states of all the agents constitute the global state, i.e.,
\begin{equation}\label{global state}
\boldsymbol{s}_{t}\triangleq\left[\{\boldsymbol{s}_{l,t}\}_{l\in\mathcal{L}},\boldsymbol{s}_{\rm B,\emph{t}}\right].
\end{equation}

Let $\mathcal{S}_{l}$ be the state space of the $l$-th IRS. $\mathcal{S}_{\rm B}$ and $\mathcal{S}$ are the BS agent state space and the global state space, respectively. We have $\boldsymbol{s}_{l,t}\in\mathcal{S}_{l}$, $\boldsymbol{s}_{\rm B,\emph{t}}\in\mathcal{S}_{\rm B}$, $\boldsymbol{s}_{t}\in\mathcal{S}$ and $\mathcal{S}\triangleq\left\{\bigcup_{l\in\mathcal{L}}\left\{\mathcal{S}_{l}\right\}\right\}\bigcup\mathcal{S}_{\rm B}$.

\subsubsection{Action}\label{action definition}
The local action of the BS is defined as the transmit beamforming vector at the $t$-th slot:
\begin{equation}\label{BS action}
\boldsymbol{a}_{\rm B,\emph{t}}\triangleq\left[\{\textbf{v}_{k,t}\}_{k\in\mathcal{K}}\right].
\end{equation}
The local action of each IRS implies the chosen bit resolution, the phase shift vector and the ON/OFF status vector at the $t$-th slot as
\begin{equation}\label{IRS action}
\boldsymbol{a}_{l,t}\triangleq\left[b_{l},\boldsymbol{\theta}_{l,t},\boldsymbol{\rho}_{l,t}\right].
\end{equation}
The optimization argument $\boldsymbol{\Phi}_{l},\forall l\in\mathcal{L}$ in~(\ref{optimization function}) is a matrix while matrix processing is more complex than vector processing. For simplicity, we assume that each IRS element reflects the signal independently without signal coupling. Therefore, the optimization variables $\boldsymbol{\Phi}_{l}$ is sparse and equivalent to the phase shift vector $\boldsymbol{\theta}_{l}=\left[\theta_{l,1},...,\theta_{l,N_{l}}\right]$.

The local action space of the $l$-th IRS agent is expressed as
\begin{equation}\label{local action space}
\mathcal{A}_{l}\triangleq\left\{(b_{l},\boldsymbol{\theta}_{l},\boldsymbol{\rho}_{l})|b_{l}\in \mathcal{B},\theta_{l,n}\in\mathcal{F}_{b_{l}},\rho_{l,n}\in \{0,1\},\forall n \in\mathcal{N}_{l}\right\}.
\end{equation}.

The local actions of all the agents make up the joint action at the $t$-th slot defined as
\begin{equation}\label{joint action}
\boldsymbol{a}_{t}\triangleq\left[\{\boldsymbol{a}_{l,t}\}_{l\in\mathcal{L}},\boldsymbol{a}_{\rm B,\emph{t}}\right].
\end{equation}

Let $\mathcal{A}_{\rm B}$, $\mathcal{A}\triangleq\left\{\bigcup_{l\in\mathcal{L}}\left\{\mathcal{A}_{l}\right\}\right\}\bigcup\mathcal{A}_{\rm B}$ denote the BS action space and the joint action space, respectively.

\subsubsection{Reward}\label{reward definition}
Our objective is to maximize system data rate under constraints of the user data rate request, the transmit power and the IRS energy storage buffer. The reward represents the optimization objective with constraints, thus, the instant reward at the $t$-th slot is defined as
\begin{equation}\label{reward}
\begin{aligned}
r_{t}\triangleq&\underbrace{\sum_{k=1}^{K}R_{k,t}}_{\text{part 1}}+\underbrace{\xi_{1}\cdot \sum_{k=1}^{K}\min\left\{R_{k,t}-R_{k,t}^{\rm req},0\right\}}_{\text{part 2}}\\&+\underbrace{\xi_{2}\cdot \min\left\{P^{\mathrm{max}}-\sum_{k=1}^{K}P_{k,t},0\right\}}_{\text{part 3}}\\&+\underbrace{\xi_{3}\cdot \sum_{l=1}^{L}\min\left\{E_{l}(t)-E_{l}^{\mathrm{min}},0\right\}}_{\text{part 4}},
\end{aligned}
\end{equation}
where the part 1 is the achievable system data rate, the part 2, the part 3 and the part 4 are penalty which are defined as the user data rate request satisfaction level, the transmit power constraint and the IRS energy consumption degree, respectively. $\xi_{1}, \xi_{2}$ and $\xi_{3}$ are the trade-off coefficients to balance the rate and the penalty. The return is defined as the cumulative discounted future reward as follows
$G_{t}\triangleq\sum_{k=0}^{\infty}\gamma^{k}r_{t+k}$, wherer $\gamma$ denotes the discount factor. Under the premise of satisfying the constraints, we aim to obtain an optimal policy that maximizes long-term expected return, which is defined as
\begin{equation}
    \mathds{E}_{\pi}\left[G_{t}|\boldsymbol{s}_{t}=\boldsymbol{s}\right].
\end{equation}
It is conspicuous to see that such a maximization is equivalent to solve the aforementioned optimization problem~(\ref{optimization function}).
Note that the optimal policy indicates the optimal BS transmit beamforming vector and the appropriate resolution, the phase shifting vector and the working status vector on each IRS.
The state value function $V(\boldsymbol{s})$ and the state-action value function $Q(\boldsymbol{s},\boldsymbol{a})$ can be given as $V(\boldsymbol{s})=\mathds{E}_{\pi}\left[G_t\mid \boldsymbol{s}_t=\boldsymbol{s}\right]$ and $Q(\boldsymbol{s},\boldsymbol{a})=\mathds{E}_{\pi}\left[G_t\mid \boldsymbol{s}_t=\boldsymbol{s},\boldsymbol{a}_t=\boldsymbol{a}\right]$, respectively. The value functions satisfy the Bellman equation~\cite{RLIntroduction}, and thus can be expressed as
\begin{equation}
\begin{aligned}
V(\boldsymbol{s})&=\sum_{\boldsymbol{a}}\pi(\boldsymbol{a}|\boldsymbol{s})\sum_{\boldsymbol{s}'}P_{\boldsymbol{s}\boldsymbol{s}'}^{\boldsymbol{a}}\left[r+\gamma V(\boldsymbol{s}')\right],\\
Q(\boldsymbol{s},\boldsymbol{a})&=\sum_{\boldsymbol{s}'}P_{\boldsymbol{s}\boldsymbol{s}'}^{\boldsymbol{a}}\left[r+\gamma \sum_{\boldsymbol{a}'}\pi(\boldsymbol{a}'|\boldsymbol{s}')Q(\boldsymbol{s}',\boldsymbol{a}')\right],
\end{aligned}
\end{equation}
where $P_{\boldsymbol{s}\boldsymbol{s}'}^{\boldsymbol{a}}\triangleq\mathrm{Pr}\left(\boldsymbol{s}_{t+1}=\boldsymbol{s}'|\boldsymbol{s}_{t}=\boldsymbol{s},\boldsymbol{a}_{t}=\boldsymbol{a}\right)$ is the state transition probability from the current state $\boldsymbol{s}$ to the next state $\boldsymbol{s}'$ given the current action $\boldsymbol{a}$, policy $\pi(\boldsymbol{a}|\boldsymbol{s})$ denotes the conditional
probability of taking action $\boldsymbol{a}$ on the state $\boldsymbol{s}$.

\subsection{Multi-Agent Q-mix Networks}
Considering the fact that the action space in~(\ref{local action space}) has high-dimensional and discrete characteristics, the training of the neural network would make the the computation resources overloaded. Moreover, value-based methods (e.g. Q-learning~\cite{RLIntroduction}) suffer from slow convergence and policy-based methods (e.g. DDPG~\cite{DBLP:conf/icml/SilverLHDWR14}) may be trapped in local optimal solutions. In addition, they are problematic to solve the discrete action space with high dimensionality. MADDPG~\cite{MADDPG} is a typical MARL algorithm, which works in terms of centralized training and distributed execution (CTDE) framework briefly. More specifically, it uses global information to update the centralized Q-value and renew each agent's policy through a distributed policy gradient method. However, it shows high computational complexity in large-scale action space. To handle with hybrid action space, the work in~\cite{1810.06394} proposed a framework called parameterized deep Q-networks (PDQN) to combine the algorithm DQN for discrete action space with the algorithm DDPG for continuous action space. More specifically, PDQN first chooses the continuous action $\boldsymbol{x}$ based on the discrete action $\boldsymbol{b}$, utilizes a neural network $Q(\boldsymbol{s},\boldsymbol{b},\boldsymbol{x}|\boldsymbol{\omega})$ parameterized by $\boldsymbol{\omega}$ to approximate the state-action value function and updates the network parameters by minimizing the loss function defined as
\begin{equation}\label{PDQN Loss}
\begin{aligned}
l(\boldsymbol{\omega})&=\mathds{E}\left[(y-Q(\boldsymbol{s},\boldsymbol{b},\boldsymbol{x}|\boldsymbol{\omega}))^{2}\right],\\
y&=r+\gamma\max_{\boldsymbol{b}'} Q(\boldsymbol{s}',\boldsymbol{b}',\boldsymbol{x}'|\boldsymbol{\omega}),
\end{aligned}
\end{equation}
where $y$ is the one-step discount reward value, $\boldsymbol{s}',\boldsymbol{b}',\boldsymbol{x}'$ are the global state, the discrete action vector and the continuous action vector in the next step, respectively. However, PDQN only serves the situation of the large-scale discrete-continuous hybrid action space, and thus \emph{cannot} be used in the systems with discrete-discrete hybrid action space. 

In our model, we have two levels of discrete actions of each IRS agent: the high-level action: the bit resolution $b_{l}$; the low-level action: the pair of the phase shift vector and the ON/OFF state vector $\boldsymbol{x}_{l}=(\boldsymbol{\theta}_{l},\boldsymbol{\rho}_{l})$. As for the $l$-th IRS, we first choose the high-level action $b_{l}$, which confirms the phase shift value set $\mathcal{F}_{b_{l}}$ and power consumption per element $\mu_{l}$, and then select the low-level action $\boldsymbol{x}_{l}$, which determines the reflection matrix $\boldsymbol{\Phi}_{l}$ and the power consumption $P_{l}^{\mathrm{IRS}}$ of the $l$-th IRS.

The high-level action space with $B$ elements is low-dimensional, whereas the low-level action space with the dimension increasing exponentially with the growth of the number of IRS elements. A common method to handle with such a situation is to use ergodic methods (e.g. the $\varepsilon$-greedy method~\cite{RLIntroduction}), but would lead to the execution complexity growing linearly with $|\mathcal{A}_{l}|$, which is quite intractable.

\begin{figure}[ht]
    \centering
    \includegraphics[width=8.5cm,height=9cm]{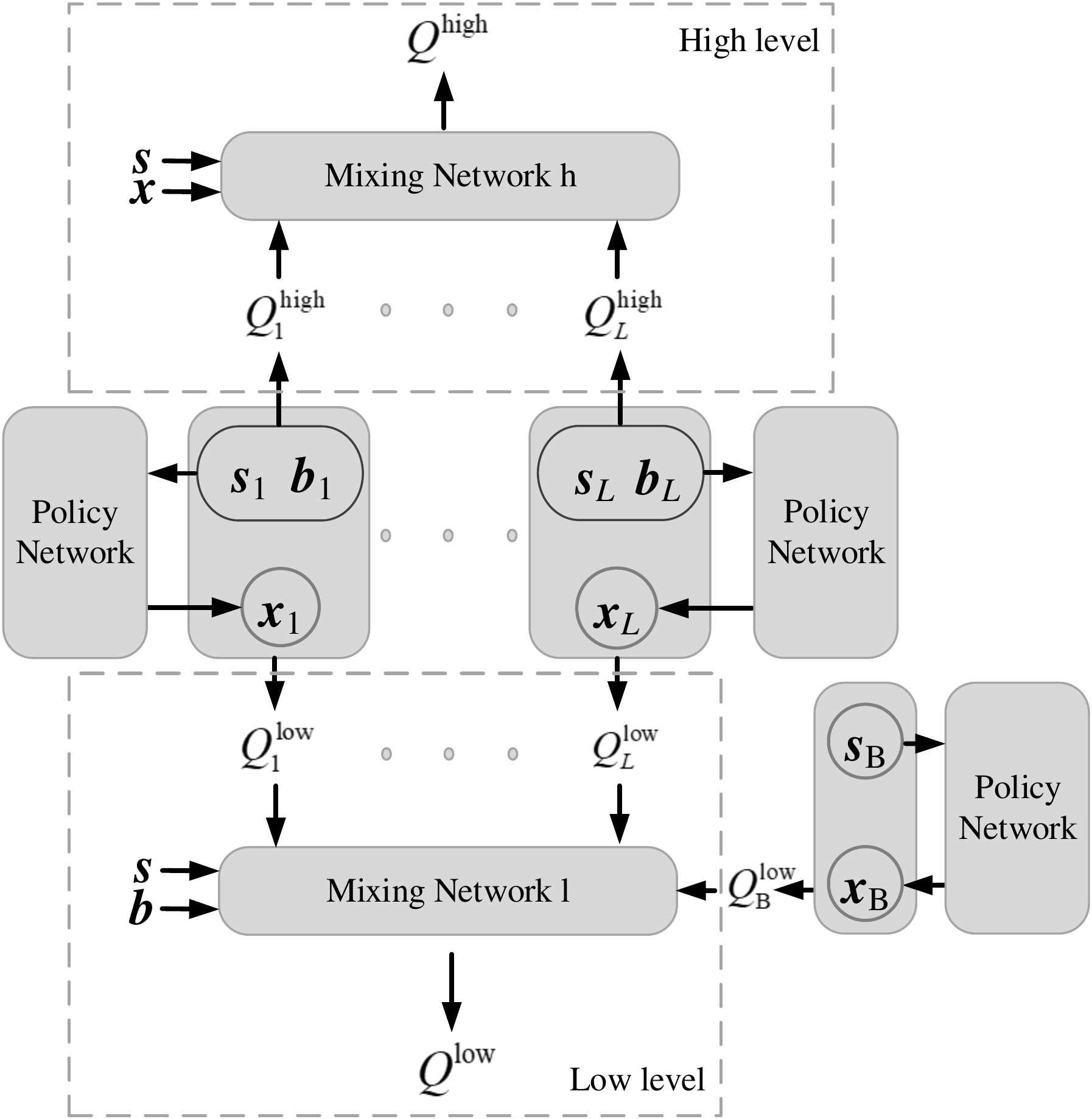}\\
    \caption{Multi-agent Q-mix framework}
    \label{MAHQMN}
\end{figure}

To cope with the aforementioned issues in this section, we propose a novel multi-agent Q-mix framework, which, as illustrated in Fig.~\ref{MAHQMN}, is made up of three parts: a high-level Q-mix network, policy networks and a low-level Q-mix network.

\subsubsection{High-level Q-mix network}\label{sec:high-level q-mix network}
It consists of $L$ agent networks and a high-level mixing network. We assume that each IRS agent employs the same structure of the agent network. The agent network performs individual value function calculations for the corresponding IRS agent. The high-level mixing network can be viewed as a monotonic function with the input of all the individual value functions.


For the IRS agent $i,\forall i\in\mathcal{L}$, the local state $\boldsymbol{s}_{l,t}$ and the bit resolution $b_{l,t-1}$ at the previous slot are utilized as the inputs of its agent network; moreover, in order to distinguish from other agents, the agent index $i$ is also utilized as the input after one-hot encoding. Using the $\varepsilon$-greedy method, with probability $\varepsilon$, the high-level action $b_{i}$ is uniformly and randomly chosen from $\mathcal{B}$; with probability $1-\varepsilon$, $b_{i}$ is selected using
\begin{equation}\label{high level action}
b_{i}=\mathop{\arg\max}_{b_{i}\in\mathcal{B}}Q_{i}^{\mathrm{high}}\left(\boldsymbol{s}_{i},b_{i}|\boldsymbol{\omega}_{i}^{\mathrm{high}}\right),
\end{equation}
where we combine the local state $\boldsymbol{s}_{i,t}$, the bit resolution $b_{i,t-1}$ and the agent number index $i$ as the new local state $\boldsymbol{s}_{i}$; the individual value function $Q_{i}^{\mathrm{high}}$ is parameterized by $\boldsymbol{\omega}_{i}^{\mathrm{high}}$. Here, we ignore the time subscript for simplicity.

The high-level mixing network at the BS takes the individual value functions of all the IRS agents as its input and the system information consisting of the global state $\textbf{s}$ and joint low-level action $\boldsymbol{x}\triangleq\left[\boldsymbol{x}_{1},...,\boldsymbol{x}_{L}\right]$ as its auxiliary message input, since the global system information is critical for the distributed agent decision. It is assumed that the global information collection and the mixing network training are executed by the BS agent. 

To guarantee high stability of the learning process and low variance of the value function, we utilize the double-network method~\cite{DQN} and construct two networks including a target network $Q^{\mathrm{high, tar}}$ parameterized by $\boldsymbol{\omega}^{\mathrm{high,tar}}$, and an evaluated network $Q^{\mathrm{high,eval}}$ parameterized by $\boldsymbol{\omega}^{\mathrm{high,eval}}$. The evaluated network evaluates the value function and updates the parameters in real time, while the target network copies the evaluated network parameters every a certain number of iterations. The parameters $\boldsymbol{\omega}^{\mathrm{ high}}=\left\{\boldsymbol{\omega}_{1}^{\mathrm{ high}},...,\boldsymbol{\omega}_{L}^{\mathrm{ high}},\boldsymbol{\omega}^{\mathrm{high,eval}}\right\}$ of the high-level network can be updated by minimizing the loss function as
\begin{equation}\label{high level loss}
\begin{aligned}
l(\boldsymbol{\omega}^{\mathrm{ high}})&\triangleq\mathds{E}\left[\left(y^{\mathrm{ high}}-Q^{\mathrm{ high,eval}}(\boldsymbol{s},\boldsymbol{b},\boldsymbol{x})\right)^{2}\right],\\
y^{\mathrm{ high}}&=r+\gamma\max_{\boldsymbol{b}'}{Q^{\mathrm{high,tar}}(\boldsymbol{s}',\boldsymbol{b}',\boldsymbol{x}')},
\end{aligned}
\end{equation}
where $y^{\mathrm{high}}$ is the one-step target value derived from the target network, $\boldsymbol{s}',\boldsymbol{b}',\boldsymbol{x}'$ are the global state, the resolution action vector and joint low-level action in the next step derived from the experience buffer, respectively.

\subsubsection{Policy network}\label{sec:policy network}
Each IRS agent $i,\forall i\in\mathcal{L}$ can obtain the low-level action $x_{i}$ through its policy network taking the IRS local state and bit resolution as input, while the BS agent can get its local action taking the BS local state as input. The IRS policy network utilizes policy methods, i.e., the Wolpertinger method and the proto-action policy gradient method, which will be discussed later, to map the continuous proto-actions with high dimensionality into actual discrete actions with low dimensionality.

\subsubsection{Low-level Q-mix network}\label{sec:low-level q-mix network}
It is made up of $L+1$ agent networks, which calculate the individual value function for all the agents, a low-level mixing network, which takes the individual value functions of all the agents as its input, and takes the global state $\boldsymbol{s}$ and the phase resolution vector $\boldsymbol{b}$ as its auxiliary information input.

Each IRS agent $i,\forall i\in\mathcal{L}$ can obtain its low-level individual value function $Q_{i}^{\mathrm{ low}}(\boldsymbol{s}_{i},b_{i},\boldsymbol{x}_{i}|\boldsymbol{\omega}_{i}^{\mathrm{low}})$ parameterized by $\boldsymbol{\omega}_{i}^{\mathrm{low}}$ and the BS agent maintain its individual value function $Q_{\mathrm{B}}^{\mathrm{low}}(\boldsymbol{s}_{\mathrm{B}},\boldsymbol{a}_{\mathrm{B}}|\boldsymbol{\omega}_{\mathrm{B}}^{\mathrm{low}})$ parameterized by $\boldsymbol{\omega}_{\mathrm{B}}^{\mathrm{low}}$. We also utilize the double-network method, that is, jointly utilize a target network $Q^{\mathrm{low,tar}}$ parameterized by $\boldsymbol{\omega}^{\mathrm{low,tar}}$ and an evaluated network $Q^{\mathrm{low,eval}}$ parameterized by $\boldsymbol{\omega}^{\mathrm{low,eval}}$. The parameters $\boldsymbol{\omega}^{\mathrm{low}}=\left\{\boldsymbol{\omega}_{1}^{\mathrm{low}},...,\boldsymbol{\omega}_{L}^{\mathrm{low}},\boldsymbol{\omega}_{\rm B}^{\mathrm{low}},\boldsymbol{\omega}^{\mathrm{low,eval}}\right\}$ of the low-level network can be updated by minimizing the loss function as
\begin{equation}\label{low level loss}
\begin{aligned}
l(\boldsymbol{\omega}^{\mathrm{low}})&\triangleq\mathds{E}\left[\left(y^{\mathrm{low}}-Q^{\mathrm{low,eval}}(\boldsymbol{s},\boldsymbol{b},\boldsymbol{x})\right)^{2}\right],\\   
y^{\mathrm{low}}&=r+\gamma\max_{\boldsymbol{x}'}{Q^{\mathrm{low,tar}}(\boldsymbol{s}',\boldsymbol{b}',\boldsymbol{x}')},
\end{aligned}
\end{equation}
where $y^{\mathrm{low}}$ is the one-step target value obtained from the low-level mixing target network.

Based on the proposed framework, we summarize a multi-agent Q-mix algorithm in Algorithm \ref{MAHQMIX algorithm}. In the training process, each agent initializes its network parameters and observes its local state, while the BS agent collects all the local states as the global state. Then, each IRS agent $i,\forall i\in\mathcal{L}$ selects the bit resolution action $b_{i}$ using the $\varepsilon$-greedy method and obtains the low-level action $\boldsymbol{x}_{i}$ through its policy network, while the BS agent gets the local action $\boldsymbol{a}_{B}$. After executing an joint action, which consists of all the local actions, all the agents receive a reward from the system and observe their next local states. Then the transition experience, which consists of the global state, the joint action, the next global state and the reward, is stored in the experience replay buffer $\mathcal{D}$. We minimize the loss function and update the parameters of all the networks through experiences sampling from the experience buffer $\mathcal{D}$. The training process is completed when the reward converges. Finally, we can obtain an optimal setting of the beamforming matrix $\textbf{V}$, the IRS bit resolution vector $\textbf{b}$, the IRS phase shift matrix $\boldsymbol{\Phi}_{i},\forall i\in\mathcal{L}$ and the ON/OFF status vector $\boldsymbol{\rho}_{i},\forall i\in\mathcal{L}$.
\begin{algorithm}
\caption{Multi-agent Q-mix}\label{MAHQMIX algorithm}
\KwIn{The channel matrix $\textbf{h}_{l,k}^{\mathrm{RU}}$,$\textbf{H}_{l}^{\mathrm{BR}}$,$\textbf{h}_{k}^{\mathrm{BU}},\forall l\in\mathcal{L},k\in\mathcal{K}$}
\KwOut{The optimal transmit beamforming matrix $\textbf{V}$, the optimal phase shift matrix $\boldsymbol{\Phi}_{l}$, the optimal ON/OFF status vector $\boldsymbol{\rho}_{l},\forall l\in\mathcal{L}$}
Initialize: The high-level Q-mix network parameters $\boldsymbol{\omega}^{\mathrm{high}}$, $\boldsymbol{\omega}^{\mathrm{high,tar}}$; The low-level Q-mix network parameters $\boldsymbol{\omega}^{\mathrm{low}}$, $\boldsymbol{\omega}^{\mathrm{low,tar}}$;  The learning rate of high-level Q-mix network,
policy network and low-level Q-mix network $\alpha_{h},\alpha_{p}$ and $\alpha_{l}$; The replay buffer $\mathcal{D}$.\;
\For{$\mathrm {epoch} = 1,2,...$$,J$}{
  initialize the state $\boldsymbol{s}_{0}$\; 
  \For{$t=0,1,\ldots,T-1$}{
    \For{$i \in \mathcal{L}$}{
      Select high-level action $b_{i}$ by the $\varepsilon$-greedy method in (\ref{high level action})\;
    }
    \For{$i \in \mathcal{L}$}{
      Get the action $\boldsymbol{x}_{i}$ through policy network\;
    }
    Get the BS action $\boldsymbol{a}_{\mathrm{B},t}$ and the IRS action $\boldsymbol{a}_{i,t},\forall i\in\mathcal{L}$\; Execute joint action $\boldsymbol{a}_{t}$ and observe the next state $\boldsymbol{s}_{t+1}$ and reward $r_{t}$\;
    Store the experience $(\boldsymbol{s}_{t},\boldsymbol{a}_{t},\boldsymbol{s}_{t+1},r_{t})$ in $\mathcal{D}$\;
    Sample a random mini-batch of $N_{\mathcal{D}}$ experiences $(\boldsymbol{s}_{i},\boldsymbol{a}_{i},\boldsymbol{s}_{i+1},r_{i})$ from $\mathcal{D}$\;
    \For{$i \in \mathcal{L}$}{
    Get the high-level $Q_{i}^{\mathrm{high}}(\boldsymbol{s}_{i},b_{i})$\;
    }
    \For{$i \in \mathcal{L}$}{
    Get the low-level $Q_{i}^{\mathrm{low}}(\boldsymbol{s}_{i},b_{i},\boldsymbol{x}_{i})$\;
    }
    Get the BS individual value function $Q_{\mathrm{B}}^{\mathrm{low}}(\boldsymbol{s}_{\mathrm{B}},\boldsymbol{a}_{\mathrm{B}})$\;
    Update the high-level Q-mix network parameters by minimizing the loss using (\ref{high level loss})\;
    Update the policy network parameters\;
    Update the low-level Q-mix network parameters by minimizing the loss using (\ref{low level loss})\;
    Update the target networks by copying parameters from the evaluated networks;\;
  }
}
\end{algorithm}

Consider the fact that each IRS policy network has a continuous output space, while each IRS agent employs a discrete action space. To solve this problem, we propose a multi-agent Q-mix with Wolpertinger policy (MAQ-WP) algorithm by utilizing the Wolpertinger policy method ~\cite{1810.06394}. We assume that each IRS agent maintains a policy network $\mu_{i}^{\boldsymbol{\xi}_{i}}(\boldsymbol{s}_{i},b_{i})$ parameterized by $\boldsymbol{\xi}_{i}$, while the BS agent manages a policy network $\mu_{\rm B}^{\boldsymbol{\xi}_{\rm B}}(\boldsymbol{s}_{\rm B})$ parameterized by $\boldsymbol{\xi}_{\rm B}$. As shown in Algorithm \ref{policy algorithm}, each $i$-th IRS agent first receives the continuous-valued proto-actions from the Wolpertinger policy given its local state and bit resolution, and then retrieves $k$ closest discrete-valued actions using the K-nearest neighbor (KNN) method~\cite{2003Machine}.
\begin{algorithm}
\caption{Wolpertinger Policy}\label{policy algorithm}
\KwIn{The local state $\boldsymbol{s}_{i}$, the high-level action $b_{i},\forall i\in\mathcal{L}$ and the Wolpertinger factor $k$}
\KwOut{The low-level action $\boldsymbol{x}_{i}=(\boldsymbol{\theta}_{i},\boldsymbol{\rho}_{i})$}
Get the proto-actions $\hat{\boldsymbol{x}}_{i}=\mu_{i}^{\boldsymbol{\xi}_{i}}(\boldsymbol{s}_{i},b_{i})$\;
Get the discrete action space $\mathcal{A}_{i,b_{i}}\subset\mathcal{A}_{i}$ given $b_{i}$\;
Use KNN to get the set of $k$ closest actions $\mathcal{A}_{i,b_{i},k}=\mathop{\arg\min}\limits_{\boldsymbol{x}_{i}\in\mathcal{A}_{i,b_{i}}}^{k}\|\boldsymbol{x}_{i}-\hat{\boldsymbol{x}}_{i}\|_{2}$\;
Get the discrete-valued low-level action $\boldsymbol{x}_{i}=\mathop{\arg\max}\limits_{\boldsymbol{x}_{i}\in\mathcal{A}_{i,b_{i},k}}Q_{i}^{\mathrm{low}}(\boldsymbol{s}_{i},b_{i},\boldsymbol{x}_{i})$
\end{algorithm}

As the low-level actions are discrete-valued, the parameters of the policy network \emph{cannot} be updated with the gradient ascent method directly. Hence, the $i$-th IRS policy network is updated approximately with continuous-valued proto-actions using the following gradient
\begin{equation}\label{IRS policy gradient}
\begin{aligned}
\nabla_{\boldsymbol{\xi}_{i}}l(\boldsymbol{\xi}_{i})&\approx\nabla_{\boldsymbol{\xi}_{i}}Q_{i}^{\mathrm{low}}\left(\boldsymbol{s}_{i},b_{i},\hat{\boldsymbol{x}}_{i}\mid\hat{\boldsymbol{x}}_{i}=\mu_{i}^{\boldsymbol{\xi}_{i}}(\boldsymbol{s}_{i},b_{i})\right)\\
&=\nabla_{\hat{\boldsymbol{x}}_{i}}Q_{i}^{\mathrm{low}}\left(\boldsymbol{s}_{i},b_{i},\mu_{i}^{\boldsymbol{\xi}_{i}}(\boldsymbol{s}_{i},b_{i})\right)\nabla_{\boldsymbol{\xi}_{i}}\mu_{i}^{\boldsymbol{\xi}_{i}}(\boldsymbol{s}_{i},b_{i}).
\end{aligned}
\end{equation}
As the BS agent employs a continuous action space, we can update the parameters of the network by the gradient ascent method directly as
\begin{equation}\label{BS policy gradient}
\begin{aligned}
\nabla_{\boldsymbol{\xi}_{\rm B}}l(\boldsymbol{\xi}_{\rm B})&=\nabla_{\boldsymbol{\xi}_{\rm B}}Q_{\rm B}^{\mathrm{low}}\left(\boldsymbol{s}_{\rm B},\boldsymbol{a}_{\rm B}\mid\boldsymbol{a}_{\rm B}=\xi_{\rm B}^{\boldsymbol{\xi}_{\rm B}}(\boldsymbol{s}_{\rm B})\right)\\
&=\nabla_{\boldsymbol{a}_{\rm B}}Q_{\rm B}^{\mathrm{low}}\left(\boldsymbol{s}_{\rm B},\mu_{\rm B}^{\boldsymbol{\xi}_{\rm B}}(\boldsymbol{s}_{\rm B})\right)\nabla_{\boldsymbol{\xi}_{\rm B}}\mu_{\rm B}^{\boldsymbol{\xi}_{\rm B}}(\boldsymbol{s}_{\rm B}).
\end{aligned}
\end{equation}

However, the proposed MAQ-WP algorithm is still computationally demanding as the search of the closest action is linear related to the Wolpertinger factor $k$ and the number of the IRS agents $L$. In addition, the approximation in~(\ref{IRS policy gradient}) may increase the variance of the individual value function $Q^{\mathrm{low}}_{i}(\boldsymbol{s}_{i},b_{i},\boldsymbol{x}_{i}),\forall i\in\mathcal{L}$. To overcome these weaknesses, we develop a multi-agent Q-mix with policy gradient (MAQ-PG) algorithm, which allows each agent to maintain an policy network $\mu_{\boldsymbol{\theta}_{i}}$ parameterized by $\boldsymbol{\theta}_{i}$ and a mapping function $\phi(\boldsymbol{e}):\mathcal{E}\rightarrow\mathcal{A}_{i,b_{i}}$ for mapping the proto-actions $\mu_{\boldsymbol{\theta}_{i}}(\boldsymbol{e}|\boldsymbol{s}_{i})$ based on a given state $\boldsymbol{s}_{i}$ to the actual low-level actions. It is assumed that, given the range set of the proto-actions $\mathcal{E}_{\boldsymbol{a}}$, the mapping function $\phi(\boldsymbol{e})$ will deterministically generate an actual low-level action, i.e., $\forall \boldsymbol{e}\in\mathcal{E}_{\boldsymbol{a}}$ such that $\phi(\boldsymbol{e})=\boldsymbol{a}$. Here, the input of the $\mu_{\boldsymbol{\theta}_{i}}(\boldsymbol{e}|\boldsymbol{s}_{i})$, $\boldsymbol{s}_{i}$, is defined by combining the $i$-th IRS agent local state and its high-level action.

In addition, the local long-term discounted function is defined as
\begin{equation}\label{local long-term discounted reward}
\begin{aligned}
    J^{\boldsymbol{\theta}_{i}}_{i}&\triangleq\mathds{E}_{\boldsymbol{s}_{i}}[V^{\boldsymbol{\theta}_{i}}_{i}(\boldsymbol{s}_{i})]=\int_{\boldsymbol{s}_{i}}d_{i}(\boldsymbol{s}_{i})V^{\boldsymbol{\theta}_{i}}_{i}(\boldsymbol{s}_{i})\, \mathrm{d}\boldsymbol{s}_{i},
\end{aligned}
\end{equation}
where $d_{i}(\boldsymbol{s}_{i})$ is the original probability distribution of the state $\boldsymbol{s}_{i}$. We use the Bellman Equation to rewrite the equation (\ref{local long-term discounted reward}) as
\begin{equation}\label{local long-term discounted reward flatten}
    J^{\boldsymbol{\theta}_{i}}_{i}=\int_{\boldsymbol{s}_{i}}d_{i}(\boldsymbol{s}_{i})\sum_{\boldsymbol{x}_{i}\in \mathcal{A}_{i,b_{i}}}\pi_{\boldsymbol{\theta}_{i}}(\boldsymbol{x}_{i}|\boldsymbol{s}_{i})Q^{\boldsymbol{\theta}_{i}}_{i}(\boldsymbol{s}_{i},\boldsymbol{x}_{i})\, \mathrm{d}\boldsymbol{s}_{i}.
\end{equation}
With a slight abuse of notation, denote the low-level action policy of the $i$-th IRS agent by $\pi_{\boldsymbol{\theta}_{i}}$. We can get the action representation form of $\pi_{\boldsymbol{\theta}_{i}}$ by accumulating the proto-action probabilities
\begin{equation}\label{action representation form}
    \pi_{\boldsymbol{\theta}_{i}}(\boldsymbol{x}_{i}|\boldsymbol{s}_{i})=\int_{\mathcal{E}_{\boldsymbol{x}_{i}}}\mu_{\boldsymbol{\theta}_{i}}(\boldsymbol{e}|\boldsymbol{s}_{i})\mathrm{d}\boldsymbol{e}.
\end{equation}
\newtheorem{lemma}{Lemma}
\begin{lemma} \label{lemma1}
Given the proto-actions $\mu_{\boldsymbol{\theta}_{i}}(\boldsymbol{e}|\boldsymbol{s}_{i})$ and the mapping function $\phi(\boldsymbol{e})$, the policy gradient can be calculated as
\end{lemma}
\begin{equation}\label{MAHQPG policy gradient}
\begin{aligned}
\nabla_{\boldsymbol{\theta}_{i}}J^{\boldsymbol{\theta}_{i}}_{i}=\mathds{E}_{\boldsymbol{s}_{i},\boldsymbol{e}}\left[\nabla_{\boldsymbol{\theta}_{i}}\log\mu_{\boldsymbol{\theta}_{i}}(\boldsymbol{e}|\boldsymbol{s}_{i})Q^{\boldsymbol{\theta}_{i}}_{i}(\boldsymbol{s}_{i},\phi(\boldsymbol{e}))\right].
\end{aligned}    
\end{equation}
\begin{IEEEproof}
See Appendix \ref{proof of lemma 1}.
\end{IEEEproof}

$\phi(\boldsymbol{e})$ can be used to obtain an action with the probability $1$, while it may not be a prior knowledge. Thus, we can construct an estimator $\hat{\phi}(\boldsymbol{x}_{i}|\boldsymbol(e)$ to approximate the action selection probability using the Kullback-Leibler (KL) divergence.
As we assume that the action $\boldsymbol{x}_{i}$ is conditionally independent with the state $\boldsymbol{s}_{i}$ given the proto-action $\boldsymbol{e}$, we denote the true probability from state $\boldsymbol{s}_{i}$ to action $\boldsymbol{x}_{i}$ by $p(\boldsymbol{x}_{i}|\boldsymbol{s}_{i})=p(\boldsymbol{e}|\boldsymbol{s}_{i})p(\boldsymbol{x}_{i}|\boldsymbol{e})$ and denote the estimator by $\hat{p}(\boldsymbol{x}_{i}|\boldsymbol{s}_{i})=p(\boldsymbol{e}|\boldsymbol{s}_{i})\hat{\phi}(\boldsymbol{x}_{i}|\boldsymbol{e})$. Thus, the KL divergence between $p$ and $\hat{p}$ is expressed as
\begin{equation}\label{KL divergence}
\begin{aligned}
    D_{\mathrm{KL}}(p||\hat{p})=\mathbb{E}_{\boldsymbol{x}_{i}}\left[\log\left(\frac{p(\boldsymbol{x}_{i}|\boldsymbol{s}_{i})}{\hat{p}(\boldsymbol{x}_{i}|\boldsymbol{s}_{i})}\right)\right]=-\mathbb{E}_{\boldsymbol{x}_{i}}\left[\log\left(\frac{\hat{\phi}(\boldsymbol{x}_{i}|\boldsymbol{s}_{i})}{p(\boldsymbol{x}_{i}|\boldsymbol{e})}\right)\right].
\end{aligned}
\end{equation}
As the denominator in~(\ref{KL divergence}) is irrelevant to the parameters of the estimator, we can neglect the denominator and update the estimator parameters by minimizing the loss function as
\begin{equation}
    l(\hat{\phi})\triangleq-\mathds{E}_{\boldsymbol{x}_{i}}\left[\log\hat{\phi}(\boldsymbol{x}_{i}|e)\right].
\end{equation}

\subsection{Computational Complexity Analysis}
The computational complexity is primarily determined by the network architectures of the high-level agent networks, the low-level agent networks, the high-level mixing network, the low-level mixing network and the policy networks. 

As for the high-level agent network, the number of neurons in the input layer is specified by the dimension of the local state and the bit resolution, which is $K+2$. The number of neurons in the output layer is 1. It is assumed that the agent network utilizes a total of $L^{\mathrm{ha}}$ fully connected neural networks, where the $l$-th ($2\leq l\leq L^{\mathrm{ha}}-1$) hidden layer contains $n_{l}^{\mathrm{ha}}$ neurons. Then, the number of the weights in the input layer, the $l$-th hidden layer, and the final hidden layer are $(K+2)n_{1}^{\mathrm{ha}}, n_{l-1}^{\mathrm{ha}}n_{l}^{\mathrm{ha}}$ and $n_{L^{\mathrm{ha}}-1}^{\mathrm{ha}}$, respectively. Similar to the high-level agent network, we assume each that low-level agent network possess $L^{\mathrm{la}}$ hidden fully connected layers, where the $l$-th ($2\leq l\leq L^{\mathrm{la}}-1$) hidden layer contains $n_{l}^{\mathrm{la}}$ neurons. Therefore, the number of the weights in the input layer, the $l$-th hidden layer, and the final hidden layer are $(K+2+2N)n_{1}^{\mathrm{la}}, n_{l-1}^{\mathrm{la}}n_{l}^{\mathrm{la}}$ and $n_{L^{\mathrm{la}}-1}^{\mathrm{la}}$, respectively. For simplicity, here we assume that each IRS agent has identical number of reflecting elements $N$. 

Moreover, we adopt $L^{\mathrm{ip}}$ hidden fully connected for IRS agent policy network and $L^{\mathrm{bp}}$ hidden fully connected for BS agent policy network, where the corresponding $l$-th ($2\leq l\leq L^{\mathrm{ip}}-1$, $2\leq l\leq L^{\mathrm{bp}}-1$) hidden layer contains $n_{l}^{\mathrm{ip}}$ and $n_{l}^{\mathrm{bp}}$ neurons, respectively. Thus for IRS policy network, the number of the weights in the input layer, the $l$-th hidden layer, and the final hidden layer are $(K+2)n_{1}^{\mathrm{ip}}, n_{l-1}^{\mathrm{ip}}n_{l}^{\mathrm{ip}}$ and $n_{L^{\mathrm{ip}}-1}^{\mathrm{ip}}$, respectively. For BS policy network, the number of the weights in the input layer, the $l$-th hidden layer, and the final hidden layer in turn are $Kn_{1}^{\mathrm{bp}}, n_{l-1}^{\mathrm{bp}}n_{l}^{\mathrm{bp}}$ and $n_{L^{\mathrm{bp}}-1}^{\mathrm{bp}}$. For high-level mixing network and low-level mixing network, the number of the total weights are given by $n^{\mathrm{hmix}}(KL+K+2L+1)(L+1)$ and $n^{\mathrm{lmix}}(KL+K+2L+2NL)(L+1)$, respectively, where $n^{\mathrm{hmix}}$ and $n^{\mathrm{lmix}}$ is the total number of the high-level and low-level mixing network neurons. 

Suppose that the computational complexity to train a single weight is $W$. Finally, the computational complexity of the proposed MAQ-WP is
$\mathcal{O}\Big(W\big[((K+2)n_{1}^{\mathrm{ha}}+\sum_{l=2}^{L^{\mathrm{ha}}-1}{n_{l-1}^{\mathrm{ha}}n_{l}^{\mathrm{ha}}}+n_{L^{\mathrm{ha}}-1}^{\mathrm{ha}})L+
((K+2+2N)n_{1}^{\mathrm{la}}+\sum_{l=2}^{L^{\mathrm{la}}-1}{n_{l-1}^{\mathrm{la}}n_{l}^{\mathrm{la}}}+n_{L^{\mathrm{la}}-1}^{\mathrm{la}})L+((K+2)n_{1}^{\mathrm{ip}}+\sum_{l=2}^{L^{\mathrm{ip}}-1}{n_{l-1}^{\mathrm{ip}}n_{l}^{\mathrm{ip}}}+n_{L^{\mathrm{ip}}-1}^{\mathrm{ip}})L+
Kn_{1}^{\mathrm{bp}}+\sum_{l=2}^{L^{\mathrm{bp}}-1}{n_{l-1}^{\mathrm{bp}}n_{l}^{\mathrm{bp}}}+n_{L^{\mathrm{bp}}-1}^{\mathrm{bp}}+
n^{\mathrm{hmix}}(KL+K+2L+1)(L+1) + n^{\mathrm{lmix}}(KL+K+2L+2NL)(L+1)\big]\Big)$, which mainly increases quadratically with the IRS number $L$ and increases linearly with the number of users $K$. 
Regarding centralized policy network, it is necessary to traverse under various resolution $\boldsymbol{b}$ to choose the joint low-level action $\boldsymbol{x}$. The computational complexity is thus grows  exponentially with the number of IRS $L$.


\section{Simulation Results}\label{sec:SIMULATION RESULTS}
In this section, we evaluate the performance of the proposed MAQ-WP and MAQ-PG algorithms numerically. In our simulations, we consider the multi-IRSs-assisted down-link communication system described in Section~\ref{sec:System Model} and assume that each agent has a full knowledge of perfect channel information. 

\subsubsection{Simulation settings}\label{sec:simulation settings}
The BS equipped with 4 antennas is located at the coordinate $(0,0)$. The users are uniformly distributed in the ring area centered on the BS with the inner radius 100\,m and the outer radius 120\,m. The IRSs are uniformly distributed on the circle with the center of the BS and the radius of 100m. In addition, the maximum BS transmit power $P^{\mathrm{max}}$ is set to be varing from 5dBm to 30dBm, and the system noise power is set to be -80dBm. We assume that the channel $\textbf{h}_{k}^{\mathrm{BU}}$ follows Rayleigh fading and the IRS-assisted channels $\textbf{H}_{l}^{\mathrm{BR}},\textbf{h}_{l,k}^{\mathrm{RU}},\forall l\in\mathcal{L}$ follow Rician fading, which can be modeled as
\begin{equation}\label{channel matrix1}
\begin{aligned}
&\textbf{H}_{l}^{\mathrm{BR}}=\mathrm{PL}_{1,l}\cdot\left(\sqrt{\frac{{\epsilon}_{1}}{1+{\epsilon}}_{1}}\textbf{a}_{N_{l}}(\vartheta_{l})\textbf{a}_{M}(\varphi)^{H}+\sqrt{\frac{1}{1+{\epsilon}_{1}}}\bar{\textbf{H}}_{l}^{\mathrm{BR}}\right),\\
&\textbf{h}_{l,k}^{\mathrm{RU}}=\mathrm{PL}_{2,l,k}\cdot\left(\sqrt{\frac{{\epsilon}_{2}}{1+{\epsilon}_{2}}}\textbf{a}_{N_{l}}(\chi_{l,k})+\sqrt{\frac{1}{1+{\epsilon}_{2}}}\bar{\textbf{h}}_{l,k}^{\mathrm{RU}}\right),
\end{aligned}
\end{equation}
where ${\epsilon}_{1},{\epsilon}_{2}$ are Rician factors, $\textbf{a}(\cdot)$ is the steering vector that is defined as $\textbf{a}_{N}(x)=\left[1,e^{-j2\pi\frac{d}{\lambda}\sin{x}},...,e^{-j2(N-1)\pi\frac{d}{\lambda}\sin{x}}\right]^{T}$, $\vartheta_{l},\varphi,\chi_{l,k}$ are angular settings, $\bar{\textbf{H}}_{l}^{\mathrm{BR}}$ and $\bar{\textbf{h}}_{l,k}^{\mathrm{RU}}$ denote the non-line-of-sight (NLOS) components whose elements both follow the symmetric complex Gaussian distribution $\mathcal{CN}(0,1)$, $\mathrm{PL}_{1,l}$ denotes the path loss between the BS and the $l$-th IRS, $\mathrm{PL}_{2,l,k}$ denotes the path loss between the $l$-th IRS and the $k$-th user. These two types of path loss are defined by
\begin{equation}\label{path loss}
\begin{aligned}
\mathrm{PL}_{1,l}&=(\mathrm{PL}_{0}-10\kappa_{l}\lg(d_{l}/d_{0})),\\
\mathrm{PL}_{2,l,k}&=(\mathrm{PL}_{0}-10\kappa_{l,k}\lg(d_{l,k}/d_{0})),
\end{aligned}
\end{equation}
where $d_{l}$ is the distance between the BS and the $l$-th IRS, $d_{l,k}$ is the distance between the $l$-th IRS and the $k$-th user. The harvested energy obeys Poisson distribution with the probability function $\mathrm{Pr}(X=k)=\frac{\varsigma^{k}}{k!}e^{-\varsigma}, k=0,1,...$, where $\varsigma=2.2$. The descriptions and values of the parameters $\mathrm{PL}_{0}$, $\kappa_{l}$, $\kappa_{l,k}$, ${\epsilon}_{1}$, ${\epsilon}_{2}$ in~(\ref{channel matrix1}) and~(\ref{path loss}) are listed in Table~\ref{system parameters settings}.
\begin{table}[ht]\caption{System Model Parameters}
\centering
\begin{tabular}{c c c}
\hline
Parameters & Description & Values\\
\hline
$\mathrm{PL}_{0}$ & The path loss at the reference distance $d_{0}=1$m & 30dB\\
$\kappa_{l}$ & the path loss factor between BS and IRS & 2\\
$\kappa_{l,k}$ & the path loss factor between user and IRS & 2.8\\
${\epsilon}_{1},{\epsilon}_{2}$ & the Rician factor & 10\\
\hline
\end{tabular}
\label{system parameters settings}
\end{table}

The network parameters settings of the proposed algorithms are summarized in Table~\ref{network parameters settings}
\begin{table}[ht]\caption{Network Parameters}
\centering
\begin{tabular}{c c c}
\hline
$J$ & Total training epochs & 20000\\
$T$ & Training steps per epoch & 100\\
$D$ & Experience replay buffer size & 100000\\
$N_{\mathcal{D}}$ & Sample mini-batch & 128\\
$\gamma$ & Discount factor & 0.99\\
$\alpha_{h}$ & Learning rate of the high-level Q-mix network & 0.0001\\
$\alpha_{p}$ & Learning rate of the policy networks & 0.0001\\
$\alpha_{l}$ & Learning rate of the low-level Q-mix network & 0.0001\\
$\varepsilon$ & $\varepsilon$-greedy policy factor & $[0.02,0.2]$\\
\hline
\end{tabular}
\label{network parameters settings}
\end{table}

\subsubsection{Comparisons with benchmarks}\label{sec:comparisons with benchmarks}
\begin{figure}
    \centering
    \includegraphics[width=8.5cm,height=6.97cm]{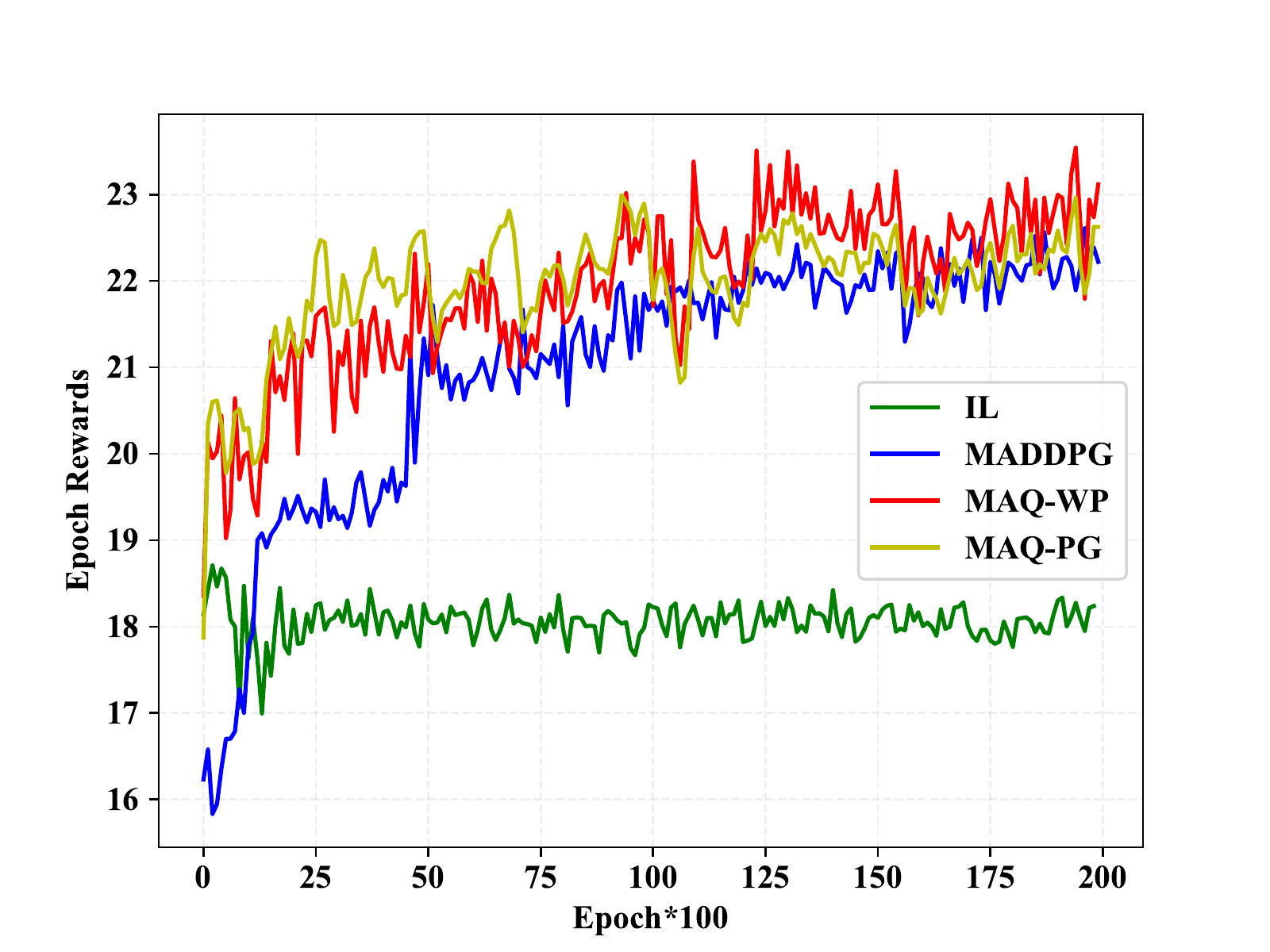}  
    \caption{Convergence comparisons of different algorithms}
    \label{fig epoch rewards}
\end{figure}

\begin{figure}
    \centering
    \includegraphics[width=8.5cm,height=6.97cm]{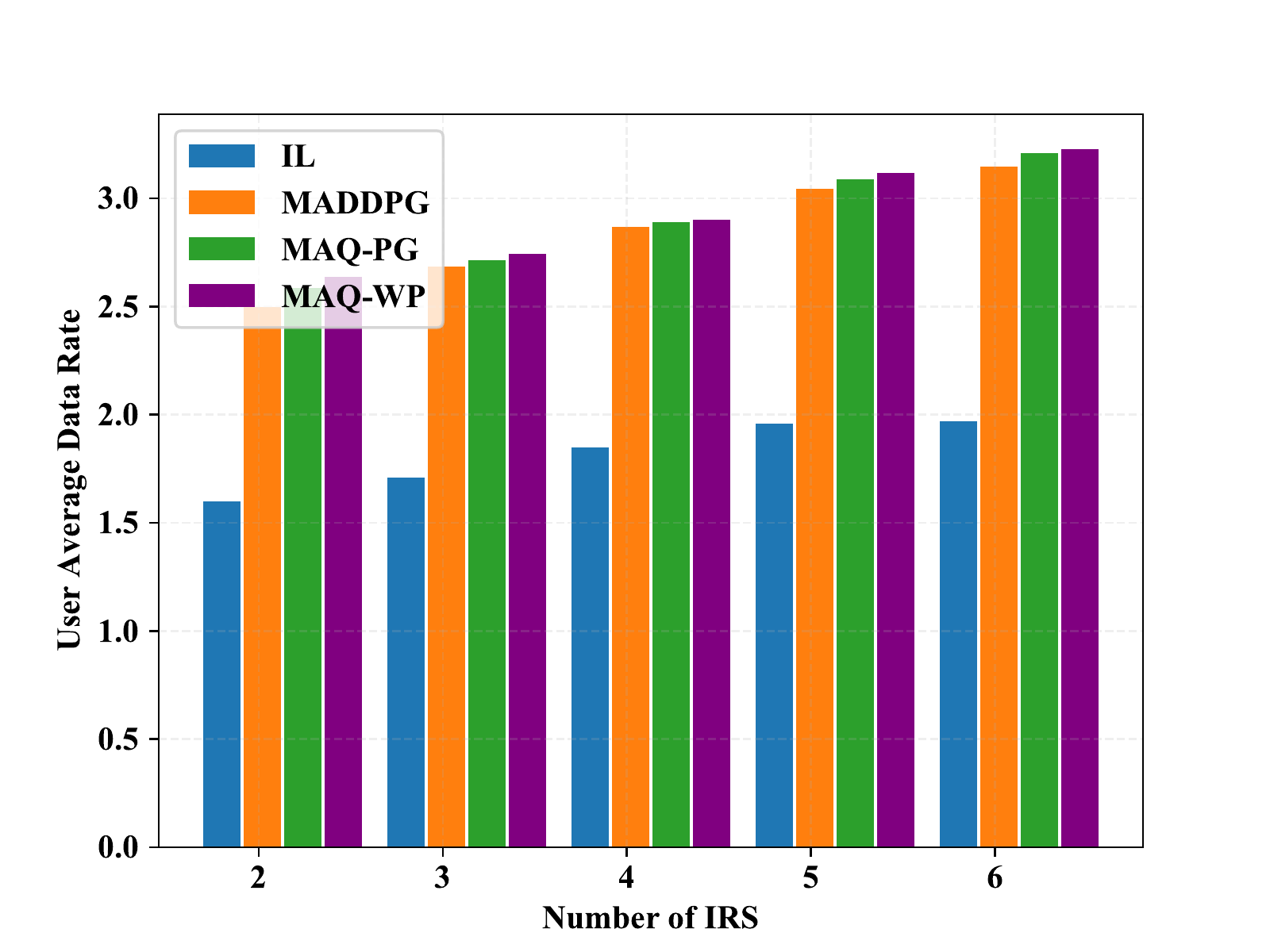}  
    \caption{User average data rate of different algorithms}
    \label{fig sum user data rate vs algorithms}
\end{figure}

We consider the number of the IRSs $L=2$, the number of the users $K=4$, the number of the elements per IRS $N_{l}=10$, the Wolpertinger policy factor $k=50$ and the maximum BS transmit power $P^{\mathrm{max}}=5$dBm. Fig.~\ref{fig epoch rewards} compares the proposed MAQ-WP and MAQ-PG algorithms with two benchmarks: MADDPG and independent learning (IL). IL means that in multi-agent environment, each agent only cares about itself and learns independently without considering the impact of other agents' actions or policies to it. It is obvious that the proposed algorithms both show appealing performance. The MAQ-WP algorithm finally achieves similar performance compared with the MAQ-PG algorithm at a larger variance as we make an approximation in~(\ref{IRS policy gradient}). The rewards of the proposed algorithms converge faster compared with MADDPG. The MAQ-PG algorithm converges at about the $5000$-th epoch and the MAQ-WP algorithm converges at about the $7500$-th epoch, while MADDPG converges at the $10000$-th epoch. This is because that our algorithms develop the hierarchical actions and calculate the gradient in~(\ref{IRS policy gradient}) and~(\ref{MAHQPG policy gradient}) based on the low-level action, while MADDPG updates the gradient by the joint action with extra computational complexity. Fig.~\ref{fig sum user data rate vs algorithms} compares the user average data rates under different algorithms with the growth of IRS number $L$. It is observed that, the average data rates increase with the number of IRS $L$, resulting from the increase in the number of beamforming policies as $L$ increases. The MAQ-PG and MAQ-WP algorithms enjoy an data rate improvement of $4.9\%\sim 8.8\%$ and $6.1\%\sim 10.7\%$ compared with MADDPG, respectively. In addition, the proposed algorithms both significantly outperform the IL. The comparisons in Figs.~\ref{fig epoch rewards}-\ref{fig sum user data rate vs algorithms} jointly indicate that our proposed algorithms enjoy both excellent network performance and fast convergence in learning. 

\subsubsection{Impact of Wolpertinger policy factor}\label{impact of Wolpertinger Policy Factor}
\begin{figure}
    \centering
    \includegraphics[width=8.5cm,height=6.97cm]{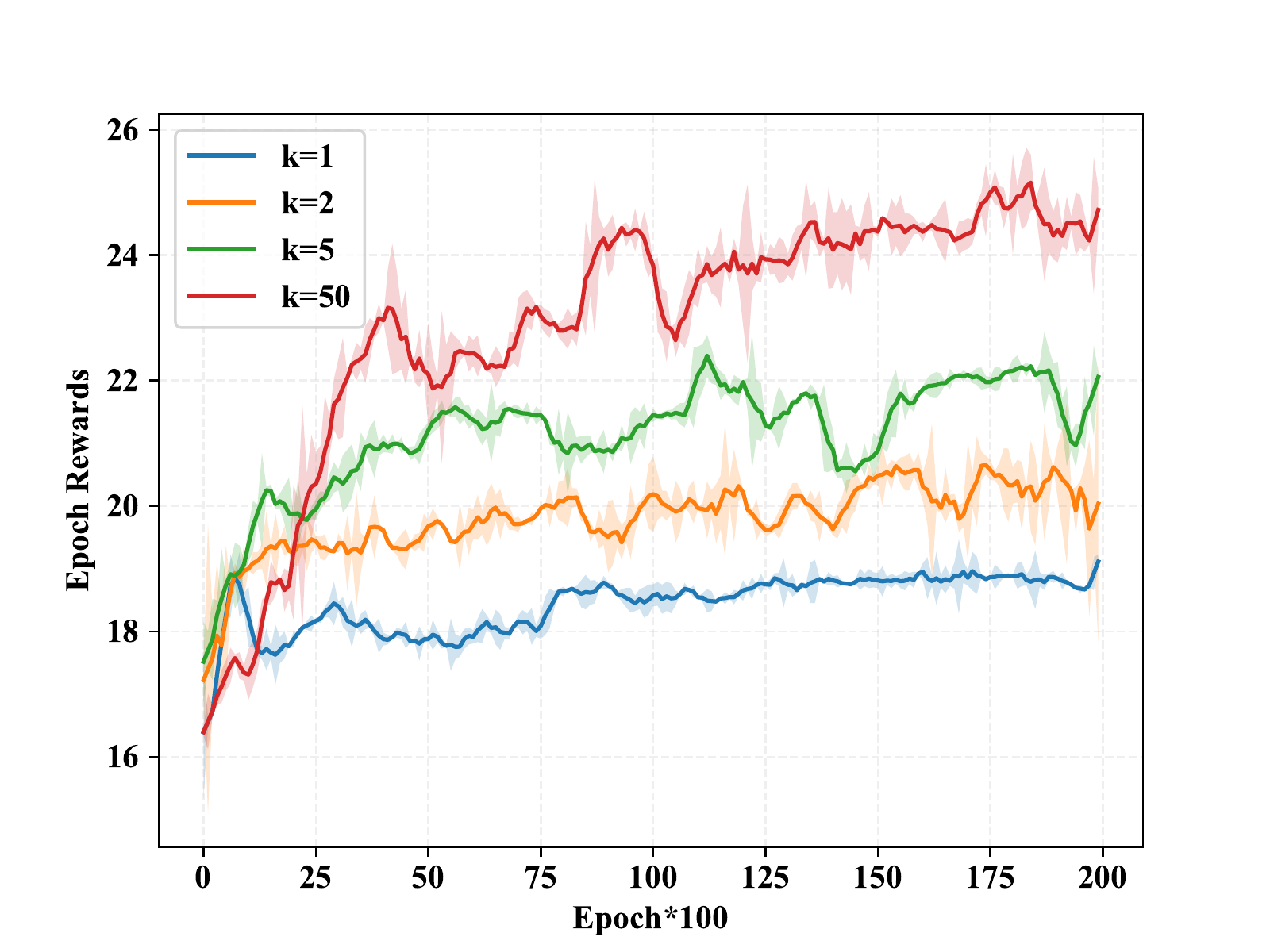}
    \caption{Performance comparisons versus epochs under different Wolpertinger policy factors}
    \label{fig Performance comparisons versus different K}
\end{figure}

Fig.~\ref{fig Performance comparisons versus different K} shows the performance of the proposed MAQ-WP algorithm under different Wolpertinger policy factors, i.e., $k=1,2,5,50$. It is worthy noting that, the solid line represents the average reward curve, and the shaded region around each learning curve shows the reward variance. It can be observed that the convergence value grows with the Wolpertinger factor $k$. The learning processes with the factor $k=1,2$ both achieve convergence after $2500$ epochs, whereas the processes with the factor $k=5,50$ converge much slower as the Wolpertinger policy selects an optimal action from $k$ actions. The results imply that if the factor $k$ is large enough, the performance can reach a considerable convergence value at cost of the training rate.

\subsubsection{Impact of maximum BS transmit power}\label{impact of maximum BS transmit power}
\begin{figure}
    \centering
    \includegraphics[width=8.5cm,height=6.97cm]{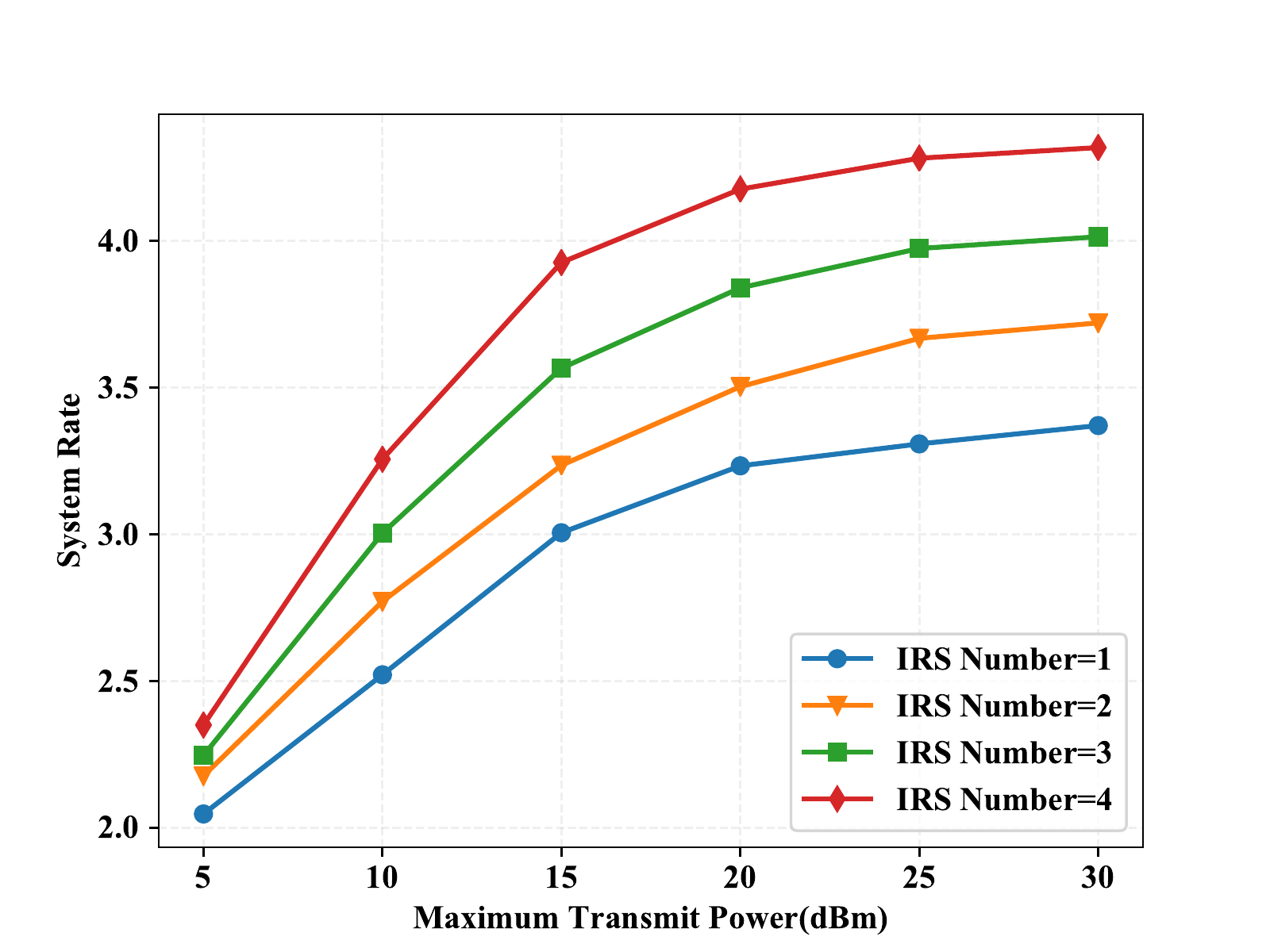}
    \caption{Performance of system rate comparisons versus different transmit power for MAQ-PG}
    \label{fig Performance comparisons versus different transmit power}
\end{figure}

Fig.~\ref{fig Performance comparisons versus different transmit power} presents the system rate of the MAQ-PG algorithm as a function of the maximum transmit power $P^{\mathrm{max}}$. It can be observed that the system rate increases steadily and tends to be stable gradually. The reason behind is that the achievable system rate increases with the transmitting signal maximum power limit. Furthermore, as the channel interferences \emph{cannot} be ignored under large $P^{\mathrm{max}}$, the system rate will eventually reach the convergence.

\begin{figure}
    \centering
    \includegraphics[width=8.5cm,height=6.97cm]{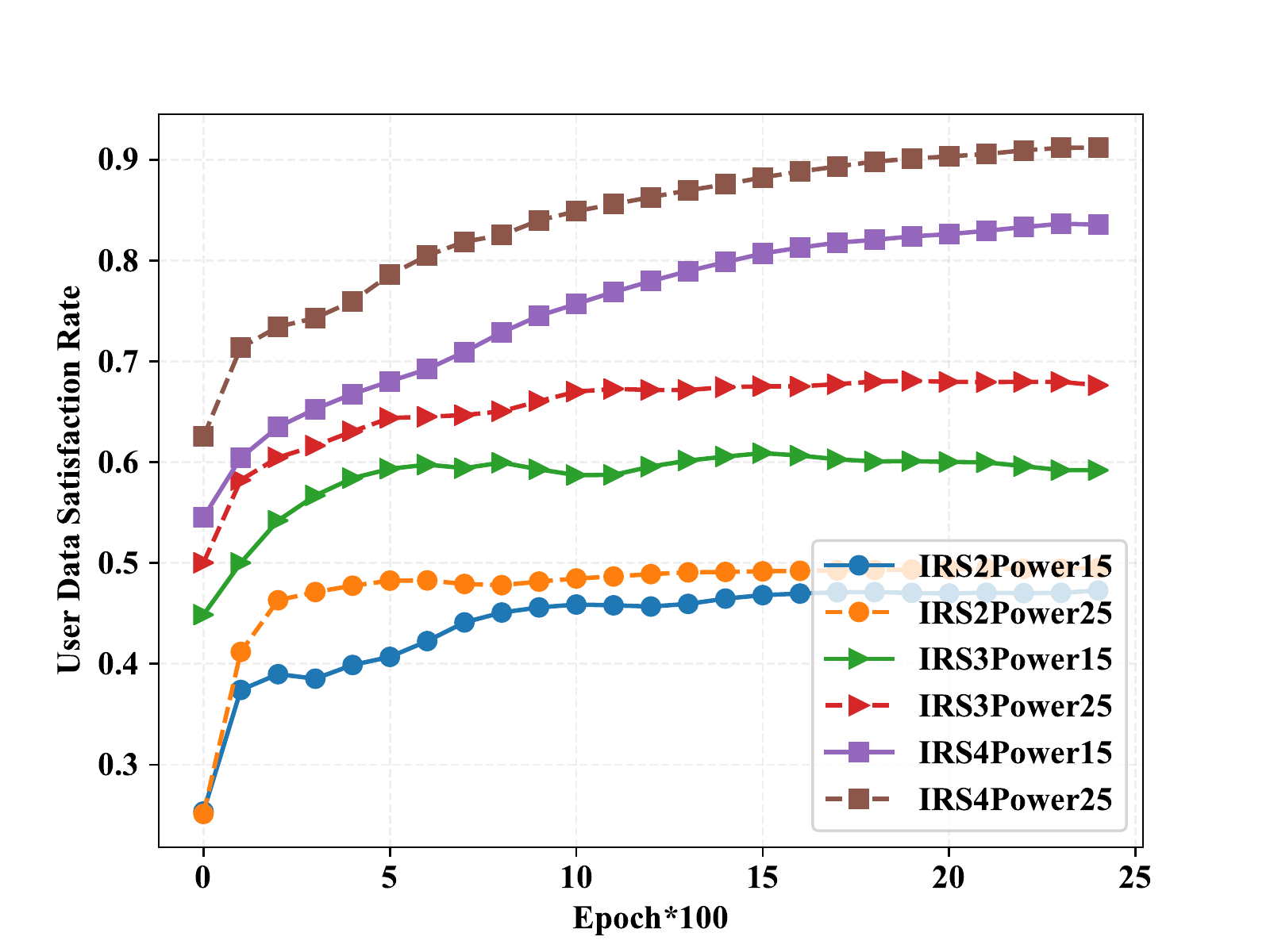}
    \caption{Performance of data satisfaction rate comparisons curve versus epochs under different transmit power for MAQ-PG}
    \label{fig Performance of data satisfactory rate curve comparisons versus different transmit power}
\end{figure}

\begin{figure}
    \centering
    \includegraphics[width=8.5cm,height=6.97cm]{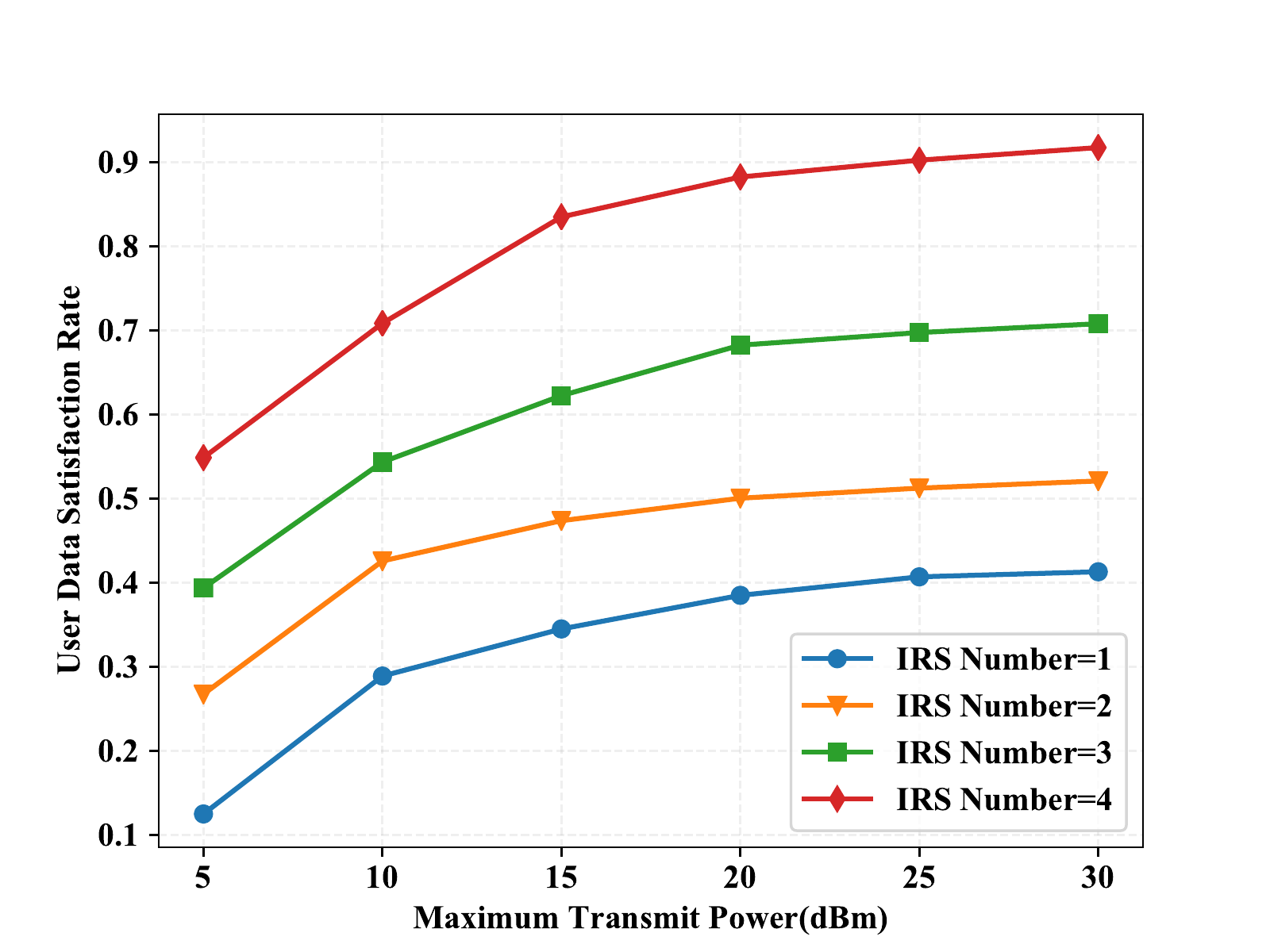}
    \caption{Performance of data satisfaction rate comparisons versus different transmit power for MAQ-PG}
    \label{fig Performance of data satisfactory rate comparisons versus different transmit power}
\end{figure}

Figs.~\ref{fig Performance of data satisfactory rate curve comparisons versus different transmit power}-\ref{fig Performance of data satisfactory rate comparisons versus different transmit power} show the user data rate satisfaction rate and the ratios of final convergence under different $P^{\mathrm{max}}$ and different numbers of IRS $L$, respectively. As expected, the user data rate satisfaction rate increases monotonically with $P^{\mathrm{max}}$ and $L$. The reason is that when $P^{\mathrm{max}}$ and $L$ increase, the received SINR in~(\ref{SINR}) increases, resulting in the improvement of the user data rate satisfaction rate.

\subsubsection{Impact of IRS element}\label{impact of IRS element}
\begin{figure}
    \centering
    \includegraphics[width=8.5cm,height=6.97cm]{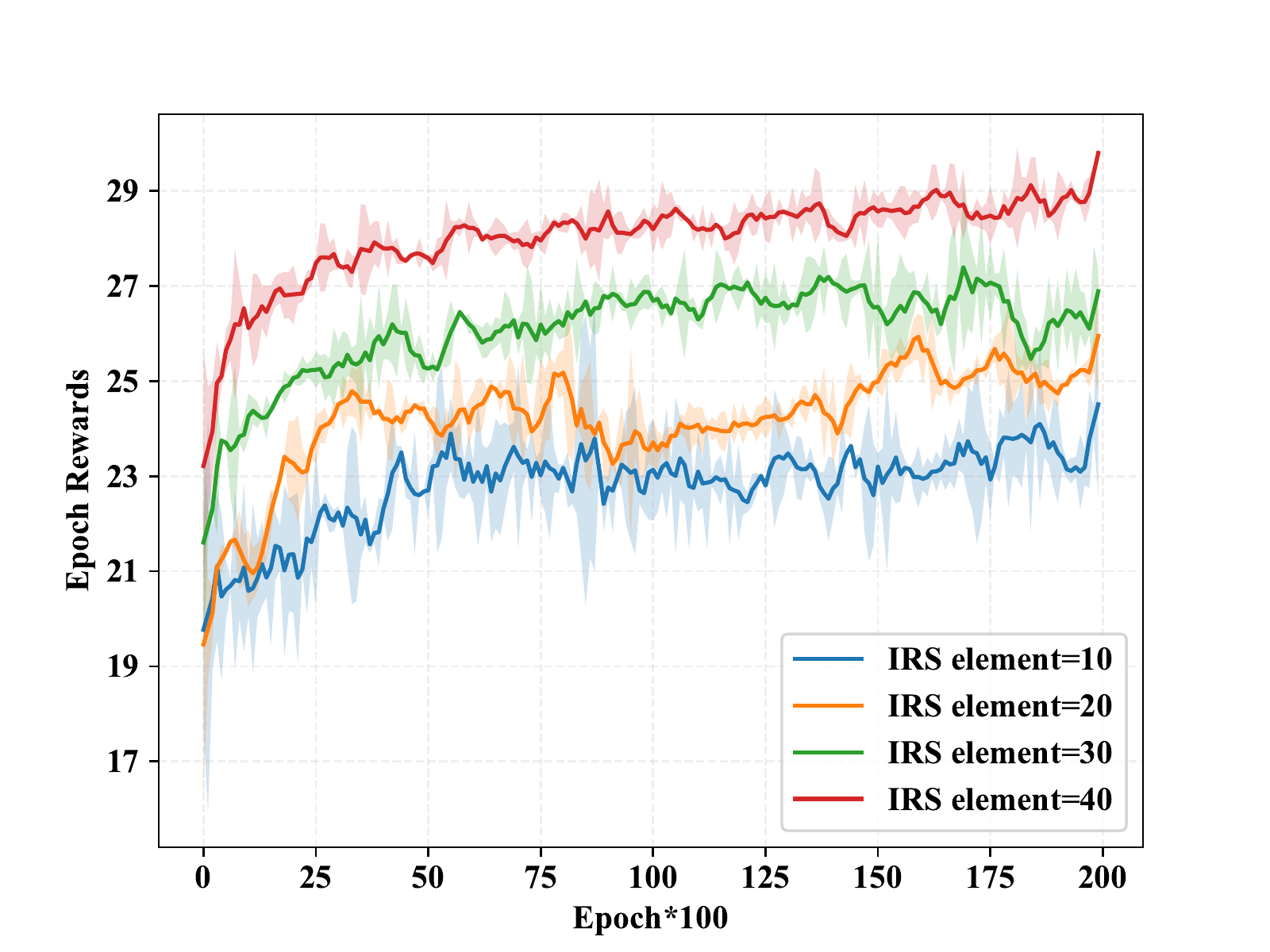}
    \caption{Performance comparisons versus epochs under different IRS elements for MAQ-PG}
    \label{fig Performance comparisons versus different IRS elements}
\end{figure}

Fig.~\ref{fig Performance comparisons versus different IRS elements} shows the epoch reward of the MAQ-PG algorithm when $N_{l}=10, 20, 30, 40$, $K=4$ and $P^{\mathrm{max}}=5$dBm. It is apparent that as $N_{l}$ increases, the convergence value of the epoch reward grows. The reason is that the IRS with more elements can be used to provide more accurate beamforming policies and more reasonable energy consumption methods, which effectively improves the instant reward and diminishes the penalties in~(\ref{reward}).

\begin{figure}
    \centering
    \includegraphics[width=8.5cm,height=6.97cm]{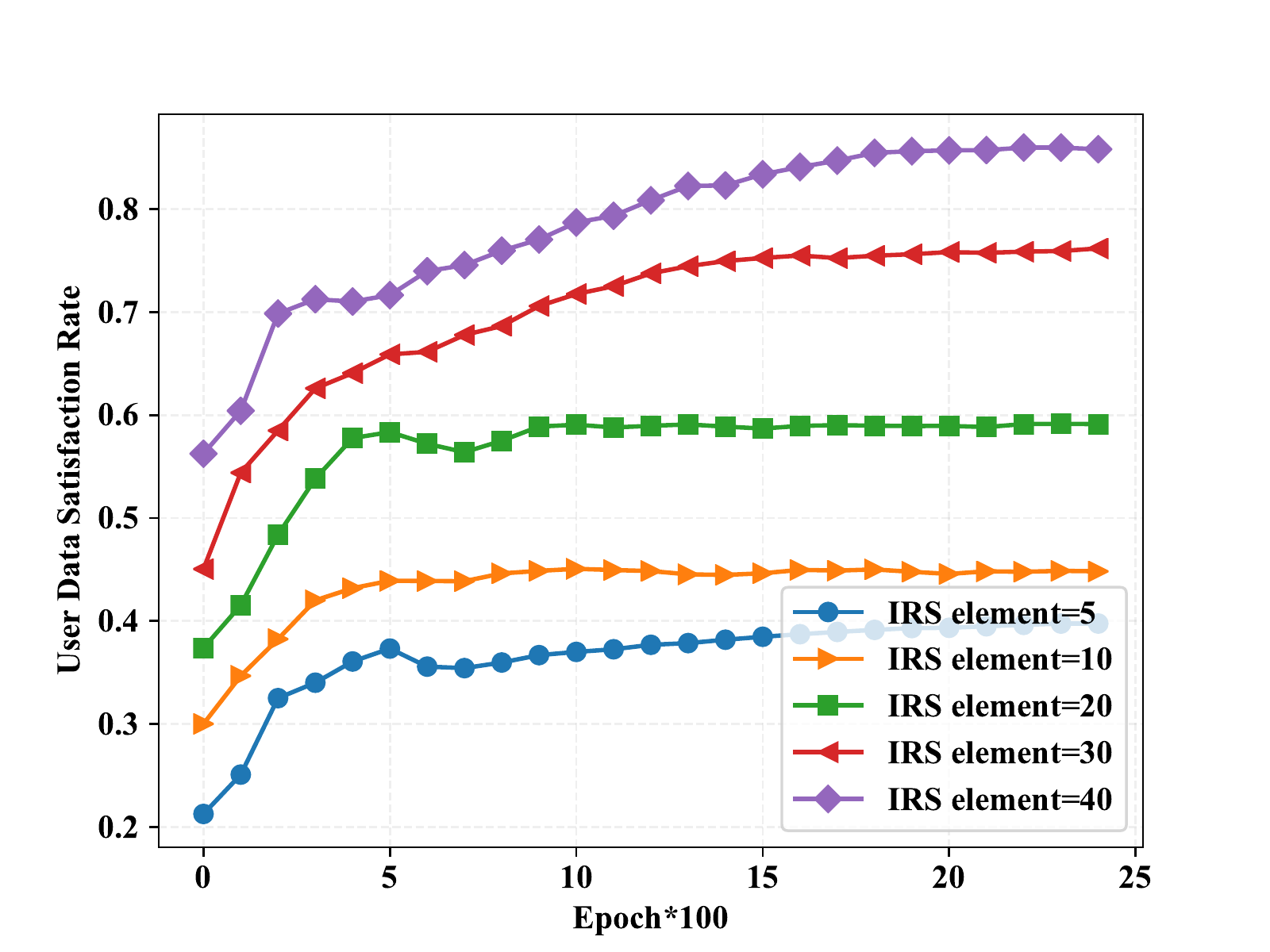}
    \caption{Performance of data satisfaction rate curve comparisons versus epochs under different IRS elements for MAQ-PG}
    \label{fig Performance of data satisfactory rate curve comparisons versus different IRS elements}
\end{figure}

\begin{figure}
    \centering
    \includegraphics[width=8.5cm,height=6.97cm]{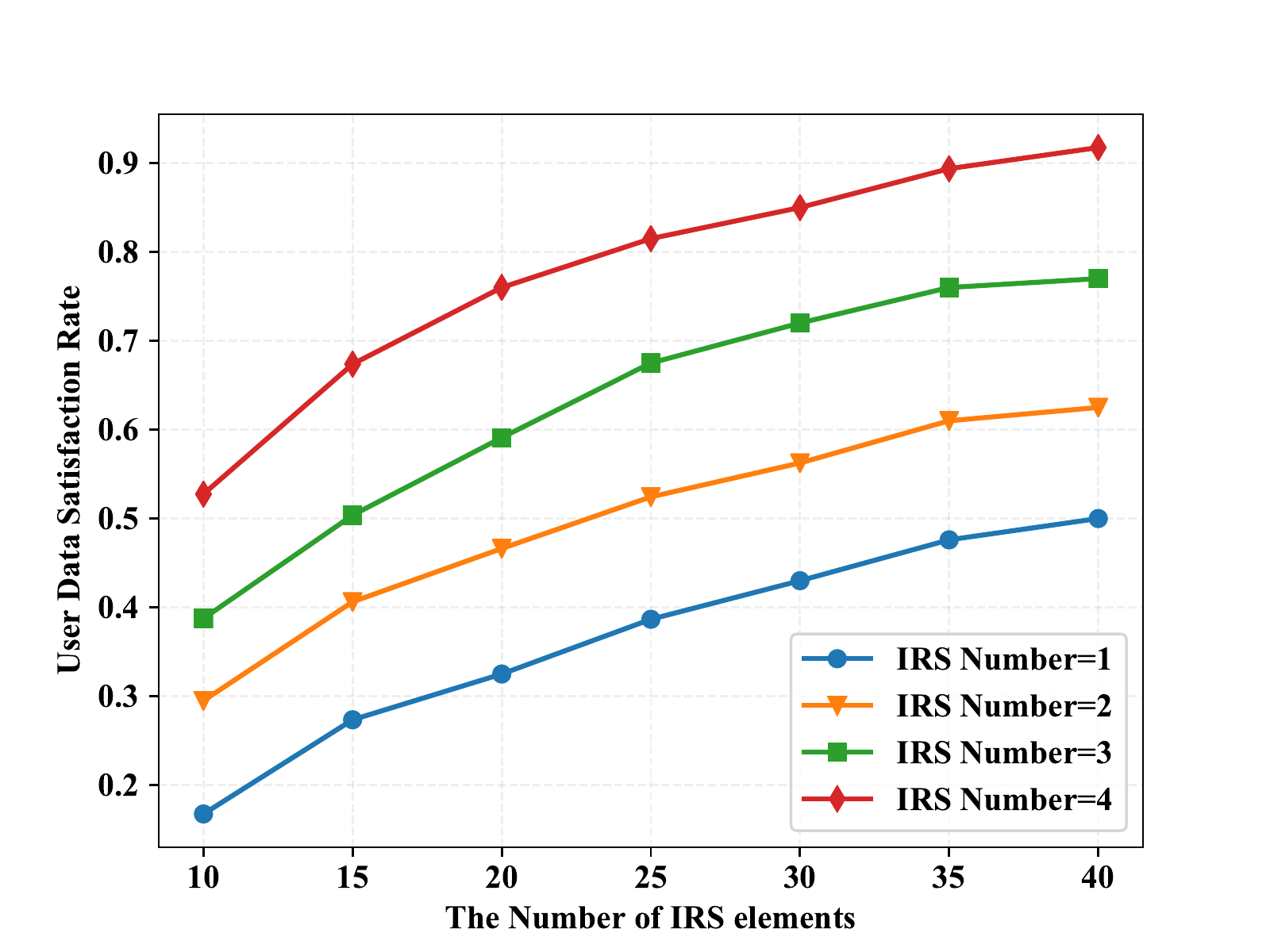}
    \caption{Performance of data satisfaction rate comparisons versus different IRS elements for MAQ-PG}
    \label{fig Performance of data satisfactory rate comparisons versus different IRS elements}
\end{figure}

Fig.~\ref{fig Performance of data satisfactory rate curve comparisons versus different IRS elements} shows the user data rate satisfaction rate of the MAQ-PG algorithm under different numbers of the IRS elements $N_{l}$. The results indicate that such a satisfaction rate increases with $N_{l}$, and achieves $91.75\%$ when $N_{l}=40$, $K=5$, $L=4$ and $P^{\mathrm{max}}=5$dBm, which implies that most of the users meet the individual requirement of their data rates. Fig.~\ref{fig Performance of data satisfactory rate comparisons versus different IRS elements} illustrates the user data rate satisfaction rate for different numbers of IRS elements $N_{l}$ and different numbers of IRSs $L$. It is also found that the user data satisfaction rate increases with $N_{l}$ and $L$. This is because that more IRSs and IRS elements, more signal paths and signal power can be reflected by the IRSs to improve the received SINR in~(\ref{SINR}). The results indicate that an appropriate $N_{l}$ is beneficial to improve the user data rate satisfaction rate; otherwise would cause the dissatisfaction of the user data rate constraints and the waste of the hardware resources.

\section{Conclusion}\label{sec:CONCLUSIONS}
In this paper, a multiple IRS-assisted multi-user communication system with an energy harvesting mechanism was considered. We formulated a BS transmit beamforming and IRSs phase shift beamforming joint optimization problem with transmit power limits, user data rate requirements and energy storage buffer constraints. We further converted this complicated non-convex optimization problem into an MDP model and designed a multi-agent Q-mix framework to decouple the optimization parameters. Moreover, the MAQ-WP and MAQ-PG algorithms were proposed to handle the high-dimensional and hybrid action space. The proposed algorithms separate the phase shift actions into two hierarchical actions, and thus are able to significantly reduce the overload of computational cost and accelerate the convergence of the learning process on the premise of a high achievable system rate. Simulation results confirmed the performance advantage of the proposed algorithms over other algorithms.



%
\bibliographystyle{IEEEtran}
\bibliography{reference}

\appendix
\subsection{Proof of \rm{Lemma 1}}\label{proof of lemma 1}
Using Lemma \ref{lemma1}, we operate the gradient of the equation (\ref{local long-term discounted reward flatten}) and unfold it as
\begin{equation}
\begin{aligned}
\nabla_{\boldsymbol{\theta}_{i}}J^{\boldsymbol{\theta}_{i}}_{i}=&\nabla_{\boldsymbol{\theta}_{i}}\int_{\boldsymbol{s}_{i}}d(\boldsymbol{s}_{i})\sum_{\boldsymbol{x}_{i}\in\mathcal{A}_{i,b_{i}}}\pi_{\boldsymbol{\theta}_{i}}(\boldsymbol{x}_{i}|\boldsymbol{s}_{i})Q_{i}^{\boldsymbol{\theta}_{i}}(\boldsymbol{s}_{i},\boldsymbol{x}_{i})\mathrm d\boldsymbol{s}_{i}\\
=&\int_{\boldsymbol{s}_{i}}d(\boldsymbol{s}_{i})\Bigg(\sum_{\boldsymbol{x}_{i}\in\mathcal{A}_{i,b_{i}}}\nabla_{\boldsymbol{\theta}_{i}}\pi_{\boldsymbol{\theta}_{i}}(\boldsymbol{x}_{i}|\boldsymbol{s}_{i})Q_{i}^{\boldsymbol{\theta}_{i}}(\boldsymbol{s}_{i},\boldsymbol{x}_{i})\\
&+\sum_{\boldsymbol{x}_{i}\in\mathcal{A}_{i,b_{i}}}\pi_{\boldsymbol{\theta}_{i}}(\boldsymbol{x}_{i}|\boldsymbol{s}_{i})\nabla_{\boldsymbol{\theta}_{i}}Q_{i}^{\boldsymbol{\theta}_{i}}(\boldsymbol{s}_{i},\boldsymbol{x}_{i})\Bigg)\mathrm d\boldsymbol{s}_{i}.
\end{aligned}
\end{equation}

Then the proof follows the procedures in~\cite{DBLP:conf/icml/SilverLHDWR14,9384286}. We can obtain
\begin{equation}
\begin{aligned}
\nabla_{\boldsymbol{\theta}_{i}}J^{\boldsymbol{\theta}_{i}}_{i}=\int_{\boldsymbol{s}_{i}}d(\boldsymbol{s}_{i})\sum_{k=0}^{\infty}\int_{\boldsymbol{s}_{i'}}\gamma^{k}p(\boldsymbol{s}_{i}\rightarrow \boldsymbol{s}_{i'},k)\\
\cdot\sum_{\boldsymbol{x}_{i}\in\mathcal{A}_{i,b_{i}}}\nabla_{\boldsymbol{\theta}_{i}}\pi_{\boldsymbol{\theta}_{i}}(\boldsymbol{x}_{i}|\boldsymbol{s}_{i})Q_{i}^{\boldsymbol{\theta}_{i}}(\boldsymbol{s}_{i},\boldsymbol{x}_{i})\mathrm d\boldsymbol{s}_{i'},
\end{aligned}
\end{equation}
where $p(\boldsymbol{s}_{i}\rightarrow \boldsymbol{s}_{i'},k)$ denotes the probability of transition from state $\boldsymbol{s}_{i}$ to state $\boldsymbol{s}_{i'}$ in $k$ steps. Let
\begin{equation}\label{stationary distribution}
\rho(\boldsymbol{s}_{i})=\sum_{k=0}^{\infty}\int_{\boldsymbol{s}_{i'}}d(\boldsymbol{s}_{i})\gamma^{k}p(\boldsymbol{s}_{i}\rightarrow \boldsymbol{s}_{i'},k)\mathrm d\boldsymbol{s}_{i'}
\end{equation}
indicate the stationary state distribution of the Markov chain starting from the state $\boldsymbol{s}_{i}$.
Substitute the equation (\ref{stationary distribution}) into the gradient and using action representation~(\ref{action representation form}) to replace the policy, resulting in
\begin{equation}
\begin{aligned}
\nabla_{\boldsymbol{\theta}_{i}}J^{\boldsymbol{\theta}_{i}}_{i}=&\int_{\boldsymbol{s}_{i}}\rho(\boldsymbol{s}_{i})\sum_{\boldsymbol{x}_{i}\in\mathcal{A}_{i,b_{i}}}\int_{\mathcal{E}_{\boldsymbol{x}_{i}}}\nabla_{\boldsymbol{\theta}_{i}}\mu_{\boldsymbol{\theta}_{i}}(\boldsymbol{e}|\boldsymbol{s}_{i})\cdot Q_{i}^{\boldsymbol{\theta}_{i}}(\boldsymbol{s}_{i},\boldsymbol{x}_{i})\mathrm d\boldsymbol{e}\mathrm d\boldsymbol{s}_{i}.
\end{aligned}
\end{equation}

In the range of $\mathcal{E}_{\boldsymbol{x}_{i}}$, each $\boldsymbol{e}$ maps a unique action $\boldsymbol{x}_{i}$, $\phi(\boldsymbol{e})=\boldsymbol{x}_{i}$. The summation of action over the action space can be substituted by the entire domain of $\boldsymbol{e}$, i.e., $\mathcal{E}$ As a result, the final gradient formula can be calculated as follows
\begin{equation}
\begin{aligned}
\nabla_{\boldsymbol{\theta}_{i}}J^{\boldsymbol{\theta}_{i}}_{i}&=\int_{\boldsymbol{s}_{i}}\rho(\boldsymbol{s}_{i})\int_{\mathcal{E}}\nabla_{\boldsymbol{\theta}_{i}}\mu_{\boldsymbol{\theta}_{i}}(\boldsymbol{e}|\boldsymbol{s}_{i})\cdot Q_{i}^{\boldsymbol{\theta}_{i}}(\boldsymbol{s}_{i},\phi(\boldsymbol{e}))\mathrm d\boldsymbol{e}\mathrm d\boldsymbol{s}_{i}\\
&=\int_{\boldsymbol{s}_{i}}\rho(\boldsymbol{s}_{i})\int_{\mathcal{E}}\mu_{\boldsymbol{\theta}_{i}}(\boldsymbol{e}|\boldsymbol{s}_{i})\\
&\cdot\nabla_{\boldsymbol{\theta}_{i}}\log\mu_{\boldsymbol{\theta}_{i}}(\boldsymbol{e}|\boldsymbol{s}_{i})\cdot Q_{i}^{\boldsymbol{\theta}_{i}}(\boldsymbol{s}_{i},\phi(\boldsymbol{e}))\mathrm d\boldsymbol{e}\mathrm d\boldsymbol{s}_{i}\\
&=\mathbb{E}_{\boldsymbol{s}_{i},\boldsymbol{e}}\left[\nabla_{\boldsymbol{\theta}_{i}}\log\mu_{\boldsymbol{\theta}_{i}}(\boldsymbol{e}|\boldsymbol{s}_{i})\cdot Q_{i}^{\boldsymbol{\theta}_{i}}\left(\boldsymbol{s}_{i},\phi(\boldsymbol{e})\right)\right].
\end{aligned}
\end{equation}




\ifCLASSOPTIONcaptionsoff
  \newpage
\fi



%

%

\begin{IEEEbiography}[{\includegraphics[width=1in,height=1.25in,clip,keepaspectratio]{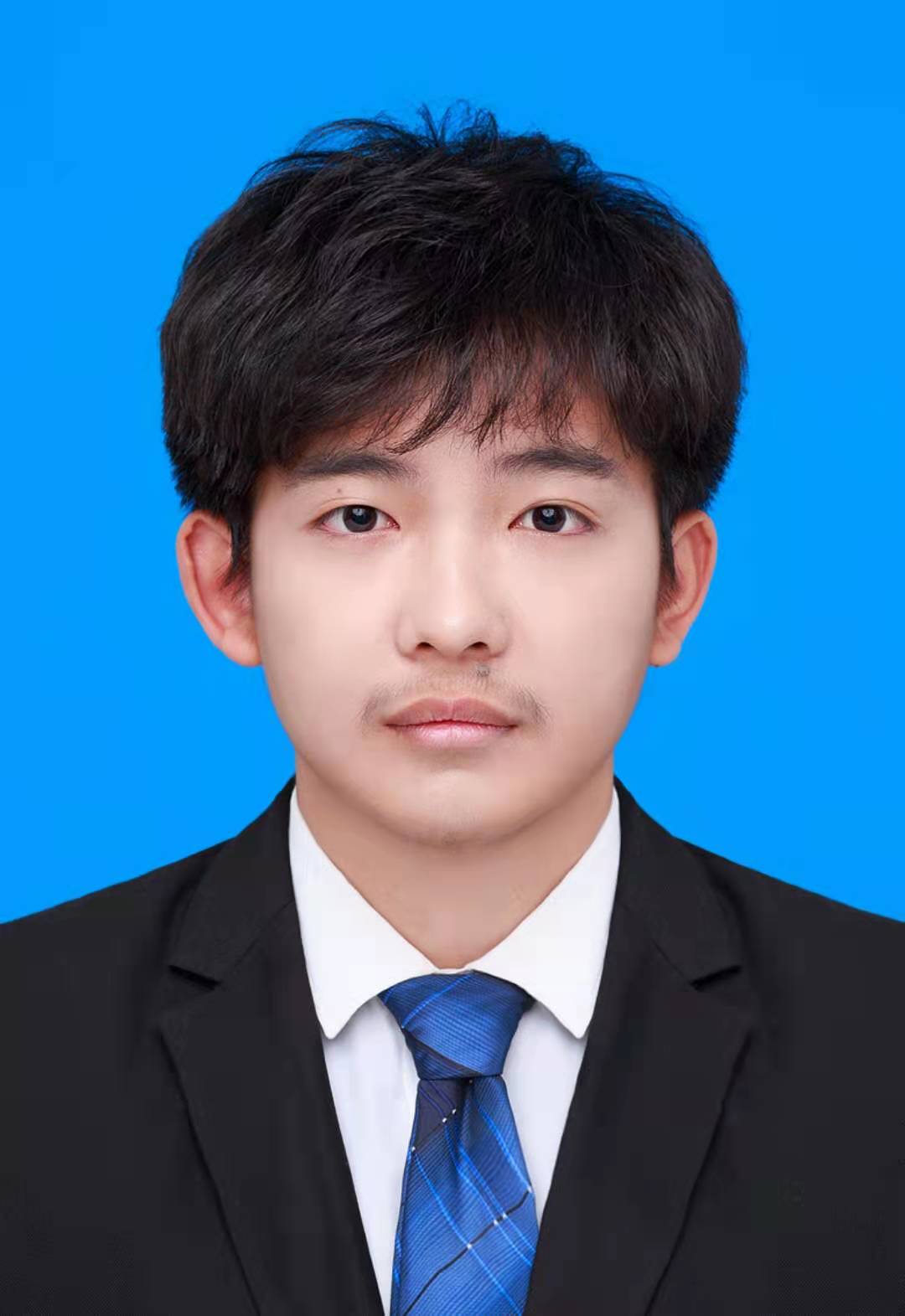}}]
{Jie Zhang}received the B.S. degree from the School of Electronic and Optical Engineering, Nanjing University of Science and Technology, Nanjing, China, in 2019, where he is pursuing the M.S. degree currently. His research interests include reinforcement learning, deep learning and intelligent reflecting surface.
\end{IEEEbiography}

\begin{IEEEbiography}[{\includegraphics[width=1in,height=1.25in,clip,keepaspectratio]{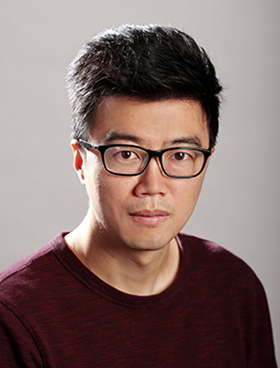}}]
{Jun Li} (M’09-SM’16) received Ph. D degree in Electronic Engineering from Shanghai Jiao Tong University, Shanghai, P. R. China in 2009. From January 2009 to June 2009, he worked in the Department of Research and Innovation, Alcatel Lucent Shanghai Bell as a Research Scientist. From June 2009 to April 2012, he was a Postdoctoral Fellow at the School of Electrical Engineering and Telecommunications, the University of New South Wales, Australia. From April 2012 to June 2015, he was a Research Fellow at the School of Electrical Engineering, the University of Sydney, Australia. From June 2015 to now, he is a Professor at the School of Electronic and Optical Engineering, Nanjing University of Science and Technology, Nanjing, China. He was a visiting professor at Princeton University from 2018 to 2019. His research interests include network information theory, game theory, distributed intelligence, multiple agent reinforcement learning, and their applications in ultra-dense wireless networks, mobile edge computing, network privacy and security, and industrial Internet of things. He has co-authored more than 200 papers in IEEE journals and conferences, and holds 1 US patents and more than 10 Chinese patents in these areas. He was serving as an editor of IEEE Communication Letters and TPC member for several flagship IEEE conferences. He received Exemplary Reviewer of IEEE Transactions on Communications in 2018, and best paper award from IEEE International Conference on 5G for Future Wireless Networks in 2017.
\end{IEEEbiography}

\begin{IEEEbiography}[{\includegraphics[width=1in,height=1.25in,clip,keepaspectratio]{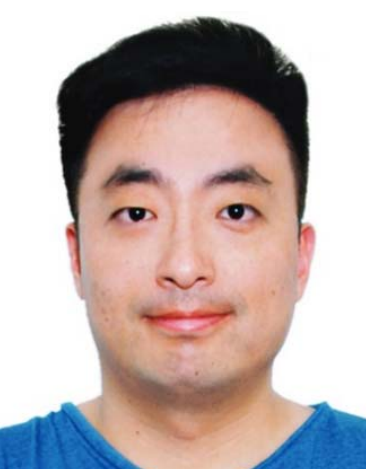}}]
{Yijin Zhang} (M’14–SM’18) received the Ph.D. degree in information engineering from The Chinese University of Hong Kong, in 2010. He joined the Nanjing University of Science and Technology, China, in 2011, where he is currently a Professor with the School of Electronic and Optical Engineering. His research interests include sequence design and resource allocation for communication networks.
\end{IEEEbiography}

\begin{IEEEbiography}[{\includegraphics[width=1in,height=1.25in,clip,keepaspectratio]{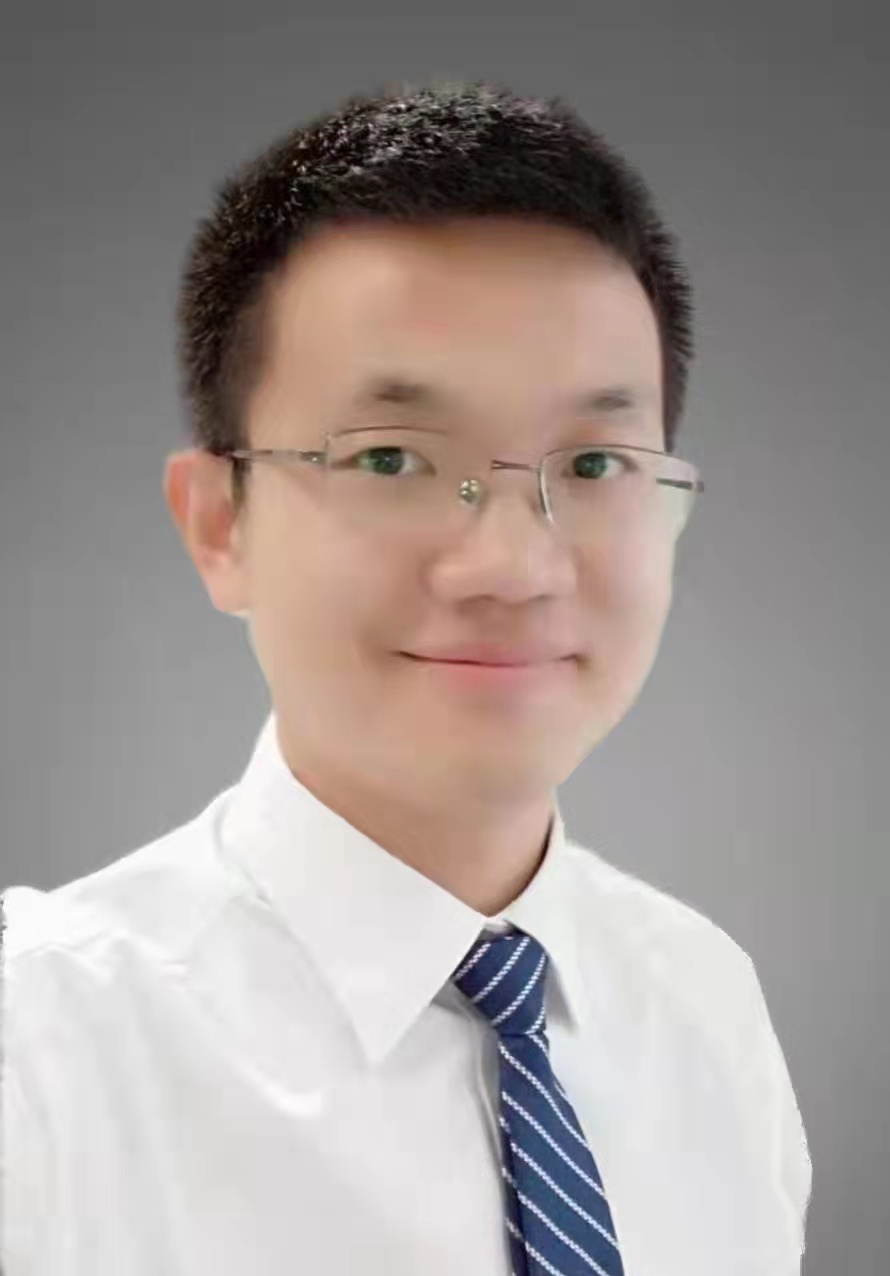}}]
{Qingqing Wu} (S’13-M’16-SM’21) received the B.Eng. and the Ph.D. degrees in Electronic Engineering from South China University of Technology and Shanghai Jiao Tong University (SJTU) in 2012
and 2016, respectively. He is currently an assistant professor with the State key laboratory of Internet of Things for Smart City, University of Macau. From 2016 to 2020, he was a Research Fellow in the Department of Electrical and Computer Engineering
at National University of Singapore. His current research interest includes intelligent reflecting surface (IRS), unmanned aerial vehicle (UAV) communications, and MIMO transceiver design. He has coauthored more than 150 IEEE papers with 24 ESI highly cited
papers and 8 ESI hot papers, which have received more than 10,000 Google citations. He was listed as Clarivate ESI Highly Cited Researcher in 2021 and World’s Top 2$\%$ Scientist by Stanford University in 2020. 

He was the recipient of the IEEE Communications Society Young Author Best Paper Award in 2021, the Outstanding Ph.D. Thesis Award of China Institute of Communications in 2017, the Outstanding Ph.D. Thesis Funding in SJTU in 2016, the IEEE ICCC Best Paper Award in 2021, and IEEE WCSP Best Paper Award in 2015. He was the Exemplary Editor of IEEE Communications Letters in 2019 and the Exemplary Reviewer of several IEEE journals. He serves as an Associate Editor for IEEE Transactions on Communications, IEEE
Communications Letters, IEEE Wireless Communications Letters, IEEE Open Journal of Communications Society (OJ-COMS), and IEEE Open Journal of Vehicular Technology (OJVT). He is the Lead Guest Editor for IEEE Journal on Selected Areas in Communications on ”UAV Communications in 5G and Beyond Networks”, and the Guest Editor for IEEE OJVT on “6G Intelligent Communications” and IEEE OJ-COMS on “Reconfigurable Intelligent Surface Based Communications for 6G Wireless Networks”. He is the workshop co-chair for IEEE ICC 2019-2022 workshop on “Integrating UAVs into 5G and Beyond”, and the workshop co-chair for IEEE GLOBECOM 2020 and ICC 2021 workshop on “Reconfigurable Intelligent Surfaces for Wireless Communication for Beyond 5G”. He serves as the Workshops and Symposia
Officer of Reconfigurable Intelligent Surfaces Emerging Technology Initiative and Research Blog Officer of Aerial Communications Emerging Technology Initiative. He is the IEEE Communications Society Young Professional Chair in Asia Pacific Region.
\end{IEEEbiography}

\begin{IEEEbiography}[{\includegraphics[width=1in,height=1.25in,clip,keepaspectratio]{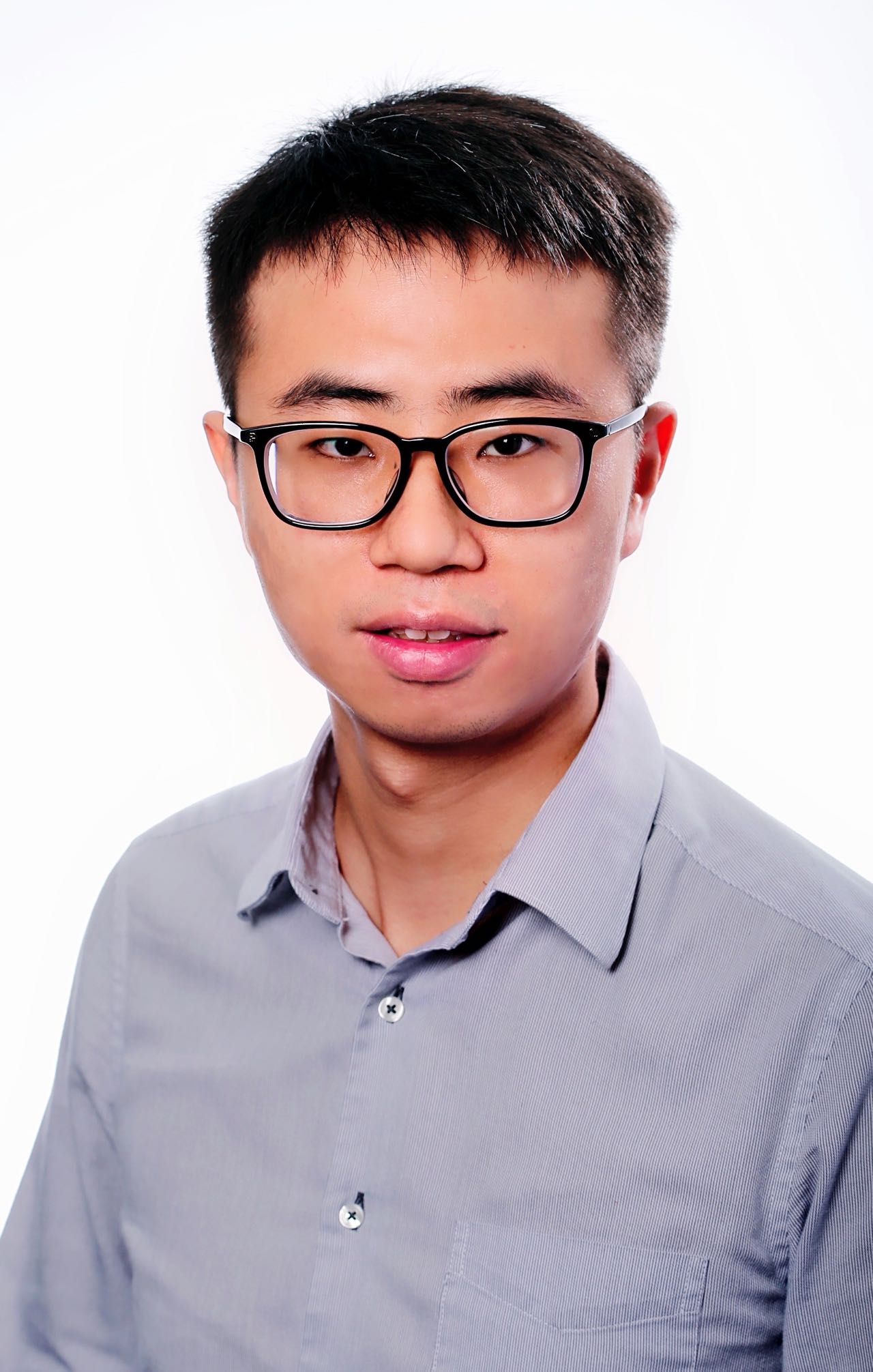}}]
{Xiongwei Wu} received the the Ph.D. degree in electronic engineering from The Chinese University of Hong Kong (CUHK), Hong Kong SAR, China, in 2021. From August 2018 to December 2018, he was a Visiting International Research Student with The University of British Columbia (UBC), Vancouver, BC, Canada. From July 2019 to January 2020, he was a Visiting Student Research
Collaborator with Princeton University, Princeton, NJ, USA. His research interests include signal processing and resource allocation, mobile edge computing, big data analytics, and machine learning.
\end{IEEEbiography}

\begin{IEEEbiography}[{\includegraphics[width=1in,height=1.6in,clip,keepaspectratio]{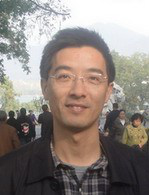}}]
{Feng Shu} (M’2016) was born in 1973. He received the Ph.D., M.S., and B.S. degrees from the Southeast University, Nanjing, in 2002, XiDian University, Xi’an, China, in 1997, and Fuyang teaching College, Fuyang, China, in 1994, respectively. From September 2009 to September 2010, he is a visiting post-doctor at the University of Texas at Dallas. From October 2005 to November 2020, he was with the School of Electronic and Optical Engineering, Nanjing University of Science and Technology, Nanjing, China, where he was promoted from associate professor to a full professor of supervising Ph.D  students in 2013. Since December 2020, he has been with the School of Information and Communication Engineering, Hainan University, Haikou, where he is currently a Professor and supervisor of Ph.D and graduate students. He is awarded with the Leading-talent Plan of Hainan Province in 2020, Fujian hundred-talent plan of Fujian Province in 2018, and Mingjian Scholar Chair Professor in 2015. His research interests include wireless networks, wireless location, and array signal processing. He was an IEEE Trans on Communications exemplary reviewer for 2020. Now, he is an editor for the journals IEEE Wireless Communications Letters and IEEE Systems Journal. He has published more than 300  in archival journals with more than 120 papers on IEEE Journals and 170 SCI-indexed papers. His citations are 4020. He holds sixteen Chinese patents and also are PI or CoPI for six national projects.
\end{IEEEbiography}

\begin{IEEEbiography}[{\includegraphics[width=1in,height=1.25in,clip,keepaspectratio]{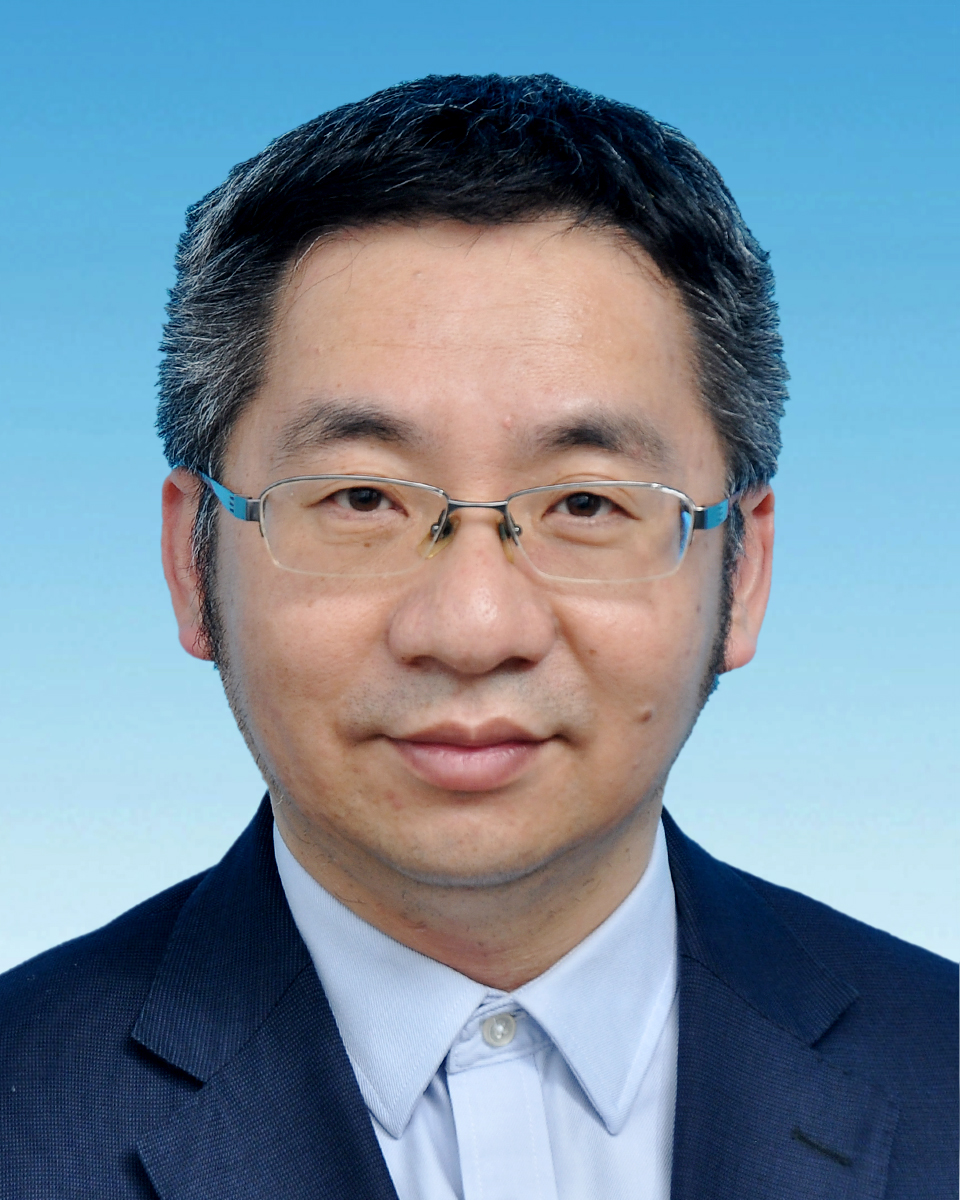}}]
{Shi Jin} received the B.S. degree in communications engineering from the Guilin University of Electronic Technology, Guilin, China, in 1996, the M.S. degree from the Nanjing University of Posts and Telecommunications, Nanjing, China, in 2003, and the Ph.D. degree in information and communications engineering from Southeast University, Nanjing, in 2007. From June 2007 to October 2009, he was a Research Fellow with the Adastral Park Research Campus, University College London, London, U.K. He is currently with the Faculty of the National Mobile Communications Research Laboratory, Southeast University. His research interests include space time wireless communications, random matrix theory, and information theory. He and his coauthors have been awarded the 2011 IEEE Communications Society Stephen O. Rice Prize Paper Award in the field of communication theory and the 2010 Young Author Best Paper Award by the IEEE Signal Processing Society. He serves as an Associate Editor for the IEEE Transactions on Communications, IEEE  Transactions on Wireless Communications, the IEEE Communications Letters, and IET Communications.
\end{IEEEbiography}

\begin{IEEEbiography}[{\includegraphics[width=1in,height=1.25in,clip,keepaspectratio]{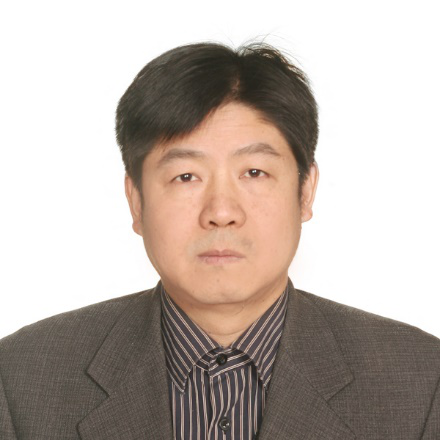}}]
{Wen Chen} (M’03–SM’11) is a tenured Professor with the Department of Electronic Engineering, Shanghai Jiao Tong University, China, where he is the director of Broadband Access Network Laboratory. He is a fellow of Chinese Institute of Electronics and the distinguished lecturers of IEEE Communications Society and IEEE Vehicular Technology Society. He is the Shanghai Chapter Chair of IEEE Vehicular Technology Society, an Editors of IEEE Transactions on Wireless Communications, IEEE Transactions on Communications, IEEE Access and IEEE Open Journal of Vehicular Technology. His research interests include multiple access, wireless AI and meta-surface communications. He has published more than 110 papers in IEEE journals and more than 120 papers in IEEE Conferences, with citations more than 7000 in google scholar.
\end{IEEEbiography}






\end{document}